\documentclass[oldversion]{aa}  
\usepackage{graphicx}
\usepackage{txfonts}
\usepackage{natbib}
\bibpunct{(}{)}{;}{a}{}{,}
\begin{document}
\title{The Environmental Dependence of Properties of Galaxies around the RDCSJ0910+54 Cluster at $z=1.1$}
\subtitle{}
\titlerunning{The Environmental Dependence of Galaxy Properties at $z=1.1$}
\authorrunning{Tanaka et al.}

\author{M. Tanaka\inst{1}, A. Finoguenov\inst{2,3}, T. Kodama\inst{4}, T. Morokuma\inst{4}, P. Rosati\inst{1}, S. A. Stanford\inst{5,6},\\P. Eisenhardt\inst{7}, B. Holden\inst{8}, S. Mei\inst{9,10}
%      \and
%      \fnmsep\thanks{}
}

\offprints{M. Tanaka}

\institute{European Southern Observatory, Karl-Schwarzschild-Str. 2
	D-85748 Garching bei M\"{u}nchen, Germany
	\email{mtanaka@eso.org}
 	\and
 	Max-Planck-Institut f\"{u}r extraterrestrische Physik, Giessenbachstrasse,
 	D-85748 Garching bei M\"{u}nchen, Germany
 	\and
 	University of Maryland, Baltimore County, 1000 Hilltop Circle,  Baltimore, MD 21250, USA
 	\and
 	National Astronomical Observatory of Japan, Mitaka, Tokyo 181-8588, Japan
 	\and
 	Institute of Geophysics and Planetary Physics, Lawrence Livermore National Laboratory L-413, 7000 East Avenue, Livermore, CA 94550, USA
        	\and
 	Department of Physics, University of California at Davis, 1 Shields Avenue, Davis, CA 95616, USA
 	\and
 	Jet Propulsion Laboratory, California Institute of Technology, Pasadena, CA 91109, USA
 	\and
 	UCO/Lick Observatories, University of California, Santa Cruz, CA 95065
	\and 
	University of Paris Denis Diderot,  75205 Paris Cedex 13, France
	\and	
	GEPI, Observatoire de Paris, Section de Meudon, 5 Place J. Janssen, 92195 Meudon Cedex, France
}

\date{Received; accepted }

\abstract{
We report on the environmental dependence of properties of galaxies around
the RDCSJ0910$+54$ cluster at $z=1.1$.
We have obtained multi-band wide-field images of the cluster with Suprime-Cam
and MOIRCS on Subaru and WFCAM on UKIRT.
Also, an intensive spectroscopic campaign has been carried out
using LRIS on Keck and FOCAS on Subaru.
We have collected 161 spectra with secure redshifts, with which
we calibrate a larger sample of photometric redshifts.
We discover a possible large-scale structure around the cluster
in the form of three clumps of galaxies.
Two out of the three newly discovered clumps of galaxies are detected
as extended X-ray sources, suggesting that they are bound systems.
There seem to be filaments of galaxies in between the clumps.
This is potentially one of the largest structures found so far in the $z>1$ Universe.
We then examine stellar populations of galaxies in the structure.
First, we quantify the color-radius relation.
Red galaxies have already become the dominant population in the cores of
rich clusters at $z\sim1$, and the fraction of red galaxies has not
strongly changed since then.
The red fraction depends on richness of clusters in the sense that
it is higher in rich clusters than in poor groups.
We confirm that this trend is not due to possible biases in photometric redshifts.
Next, we examine red sequence galaxies.
The luminosity function of red galaxies in rich clusters is
consistent with that in local clusters.
On the other hand, luminosity function of red galaxies in
poor groups shows a deficit of faint red galaxies.
This confirms our earlier findings that galaxies follow
an environment-dependent down-sizing evolution.
Interestingly, the truncation magnitude of the red sequence 
appears brighter than that found in the RDCS~J1252$-$29 field at $z=1.24$.
This suggests that there is a large variation in the evolutionary phases of
galaxies in groups with similar masses.
Further studies of high redshift clusters will be a promising way of
addressing the role of nature and nurture effects in shaping
the environmental dependence of galaxy properties observed in the local Universe.
}{}{}{}{}
% 5 {} token are mandatory
 
%\abstract{}


\maketitle

%-------------------------------------------------------
\section{Introduction}

While galaxy clusters at $z>1$ are now within our reach, larger scale 
structures are still difficult to locate and characterize. 
There are a handful of $z > 1$ clusters known to date
and detailed analyses of their cluster galaxies have been made
(e.g., \citealt{blakeslee03,lidman04,rosati04,nakata05,stanford05,stanford06,mei06a,mei06b,demarco07,tanaka07a}).
There is clear evidence that red early-type galaxies have become
a dominant population by $z=1$ in rich clusters.
A tight red sequence has already been in place (e.g., \citealt{blakeslee03,lidman04})
and spectra of $z\sim1$ cluster galaxies do not typically show signs of
active star formation \citep{mullis05,demarco07}. 
The majority of cluster galaxies seem to have become red and dead by $z=1$.

While rich clusters have been intensively studied,
poor groups at $z>1$ remain largely unexplored.
This has been hampered by the fact that high redshift poor groups are
difficult to find.  But the recent advent of large telescopes
with wide-field capabilities along with deep X-ray surveys have pushed
studies of poor groups beyond a redshift of unity.
These high redshift poor groups are particularly interesting
since they should be the precursors of present-day clusters with moderate masses.
Furthermore, there is
some observational evidence that galaxy properties depend
on cluster richness in the sense that cluster galaxies are
more evolved than group galaxies
\citep{zabludoff98,tran01,merchan02,balogh02,martinez02,tanaka05,tanaka07a,koyama07,gilbank08}.
Galaxies in rich clusters and in poor groups may follow
different evolutionary paths.
Therefore, it is essential to observe the full range of local galaxy density 
at various redshifts up to and beyond $z \sim 1$.

We are undertaking a high redshift cluster survey with
the Subaru telescope \citep{kodama05}, taking advantage of its wide-field
imaging capabilities.
Our aim is to observe a range of environments at various redshifts
and improve our understanding of galaxy evolution as a function of environment.
As part of this survey,
we have observed several high redshift clusters 
\citep{kodama05,nakata05,umetsu05,tanaka05,tanaka06,tanaka07a,tanaka07b,koyama07}.
In this paper, we focus on one of the $z>1$ clusters from our sample,
RDCSJ0910+54, which was discovered in the ROSAT Deep Cluster Survey \citep{rosati98}.
\citet{stanford02} carried out spectroscopic follow-up observations of
this cluster, which was confirmed to lie at $z=1.1$, and presented the results of a 
Chandra observation which measured a relatively high X-ray temperature. 
\citet{mei06a} presented a detailed analysis of the red sequence galaxies 
using high resolution ACS images.
Following these studies, we have carried out deep and wide-field observations of
this cluster and we present combined analyses of the photometric and spectroscopic
properties of galaxies over a wider field around the central cluster at $z=1.1$.

The layout of this paper is as follows.
We summarize our observations in Section 2.
Section 3 describes photometric catalogs and photometric redshifts.
Then, we present large-scale structures around the RDCSJ0910 cluster in Section 4.
We carry out detailed analyses of the photometric and spectroscopic properties
of galaxies in the structures in Sections 5 and 6.
Section 7 discusses implications of our results for galaxy evolution,
and the paper is summarized in Section 8.

Unless otherwise stated, we adopt H$_0=70\rm km\ s^{-1}\ Mpc^{-1}$,
$\Omega_{\rm M}=0.3$, and $\Omega_\Lambda =0.7$.
Magnitudes are on the AB system.
We use the following abbreviations: CMD for color-magnitude diagram,
and LF for luminosity function.

%-------------------------------------------------------
\section{Observations}
\subsection{Imaging Observations}

\begin{figure}
\centering
\includegraphics[width=8cm]{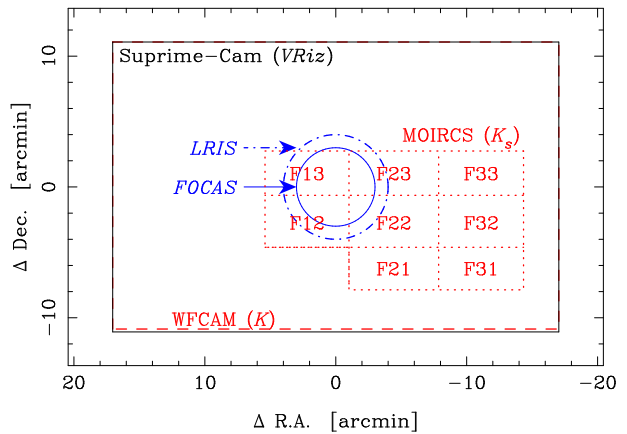}
\caption{
The area of the sky covered by our data.
The RDCSJ0910 cluster lies at ($\Delta$R.A., $\Delta$Dec.)=(0\arcmin,0\arcmin).
The solid, dashed, and dotted rectangles show the field coverages of
Suprime-Cam ($VRiz$), WFCAM ($K$), and MOIRCS ($K_s$) images, respectively.
We made an 8-pointing observation with MOIRCS as indicated by F12 -- F33.
The solid circle shows the FOCAS field.
We have made several pointings with LRIS and the dot-dashed circle
shows an approximate field coverage of the LRIS observations.
}
\label{fig:data}
\end{figure}

We made Suprime-Cam observations in $R$ and $z$ bands in May 2005.
Following this initial run, we carried out $VRiz$ bands
observations with Suprime-Cam in November 2006,
$K_s$-band observation with MOIRCS in February 2006,
and $K$-band observation with WFCAM in March 2007.
Most of the observations were carried out under good conditions.
We follow a standard data reduction scheme with custom designed pipelines \citep{yagi02}.
The PSF sizes vary from band to band and we smooth them to a common seeing
as detailed below.
The photometric zero-points were derived from standard star observations
for the optical images and from the 2MASS catalog \citep{jarrett00} for the nearIR images.
The Galactic extinction is corrected using the extinction map provided
by \citet{schlegel98}.
Fig. \ref{fig:data} shows the sky area covered by our data and
Table \ref{tab:data} summarizes the exposure times and limiting magnitudes.

%-------------------------------

%-------------------------------
\begin{table}
\caption{Summary of the imaging data.  The magnitude limits are $5\sigma$
limits in 2\arcsec diameter apertures.}
\label{tab:data}
\centering
\begin{tabular}{c c c}
\hline\hline
Observation         & Exposure Time & Magnitude Limit \\
\hline
Suprime-Cam: $V$    & 60min         & 26.6\\
Suprime-Cam: $R$    & 90min         & 26.7\\
Suprime-Cam: $i$    & 45min         & 26.2\\
Suprime-Cam: $z$    & 51min         & 25.2\\
WFCAM: $K$          & 42min         & 22.3\\
MOIRCS: $K_s$ - F12 & 30min         & 23.3\\
MOIRCS: $K_s$ - F13 & 33min         & 23.3\\
MOIRCS: $K_s$ - F21 & 23min         & 23.1\\
MOIRCS: $K_s$ - F22 & 33min         & 23.3\\
MOIRCS: $K_s$ - F23 & 15min         & 22.9\\
MOIRCS: $K_s$ - F31 & 30min         & 23.3\\
MOIRCS: $K_s$ - F32 & 33min         & 23.3\\
MOIRCS: $K_s$ - F33 & 33min         & 23.3\\
\hline
\end{tabular}
\end{table}

%-------------------------------------------------------
\subsection{Spectroscopic Observations}
Follow-up spectroscopic observations have been carried out
with LRIS on Keck and FOCAS on Subaru.
The LRIS observations are summarized in \citet{mei06a}.
Objects targeted for the observations are $K$-band selected
($K<20.0$ on the Vega system),
with a supplimental $i$-band selection to fill the masks.
We did not give priorities to galaxies on the red sequence.

The FOCAS observations were made in April 2006 and December 2006 in MOS mode.
We used the 300R grism with the SO58 order-cut filter.
The slit width was set to 0.8\arcsec, which gave a resolution of $R\sim500$.
In the slit assignment, we gave the highest priority to bright
($z<22.5$) galaxies on the red sequence in the $R-z$ color.
The second priority was given to fainter ($22.5<z<23.5$)
red sequence galaxies.
We manufactured two masks both centered at the RDCSJ0910 cluster.
One mask was exposed for 140min and the other for 160min.
Observing conditions were good and the seeing varied between 0.7 and 1\arcsec.

The data reduction is performed in a standard manner.
Fluxes are calibrated with standard stars observed each night.
Telluric extinctions are not corrected.
The Galactic extinction is corrected assuming the extinction
curve derived by \citet{cardelli89}.
We have visually inspected all the spectra and assign redshifts and
confidence flags with custom designed software.

We obtain 161 secure redshifts out of 237 galaxies observed in total.
We estimate possible redshifts for 22 galaxies, and
the rest of the galaxies have no secure/possible redshifts.
As we discuss later, we use photometric redshifts to extract
galaxies around the cluster redshift.
32 galaxies at $0.92\leq z_{phot}\leq 1.12$ are observed, out of which
we obtain 22 secure, 4 possible, and 6 no redshifts.
Only one of the 22 secure redshifts lies outside of the photo-$z$ range.
Three objects are observed with both LRIS and FOCAS,
only one of which has secure redshifts: LRIS-1101 and FOCAS-2-11.
Its redshifts from the LRIS and FOCAS observations are
$1.0988_{-0.0003}^{+0.0003}$ and $1.0991_{-0.0018}^{+0.0022}$,
being fully consistent within the error.
We present in Table \ref{tab:spec_cat} the catalog of the spectroscopic objects.
This includes 161 secure redshifts and 22 possible redshifts.

%-------------------------------
\begin{table*}
\caption{Spectroscopic catalog.
The columns show object ID, coordinates in J2000, redshift, confidence flag
(0 and 1 mean secure and possible redshifts, respectively),
and aperture magnitudes within 2\arcsec apertures and their errors.
The systematic zero point errors are not included here.
The redshift error does not include wavelength calibration error which is
typically $0.3\rm\AA$.
{\it The table will appear in its entirety in the electric edition of the journal.}}
\label{tab:spec_cat}
\centering
\begin{tiny}
\begin{tabular}{l l l l l l r l r l r l r l r}
\hline\hline
ID & R.A.  & Dec.  & redshift  & flag & $V$ & $\sigma(V)$ &  $R$ & $\sigma(R)$ & $i$ & $\sigma(i)$ & $z$ & $\sigma(z)$ & $K$ & $\sigma(K)$ \\ 
\hline
LRIS-8 &        09 10 47.5 & 54 17 04 & $0.6122^{+0.0002}_{-0.0002}$ & 1 &     23.65 & 0.02 & 22.83 & 0.01 & 22.43 & 0.01 & 22.19 & 0.01 & 21.14 & 0.08 \\
LRIS-9 &        09 10 51.7 & 54 22 18 & $0.0000^{+0.0000}_{-0.0000}$ & 0 &     20.87 & $<0.01$ & 20.02 & $<0.01$ & 19.44 & $<0.01$ & 19.11 & $<0.01$ & 17.91 & 0.01 \\
LRIS-11 &       09 10 38.2 & 54 22 39 & $0.0000^{+0.0000}_{-0.0000}$ & 0 &     23.00 & 0.01 & 21.93 & $<0.01$ & 20.27 & $<0.01$ & 19.57 & $<0.01$ & 18.55 & 0.01 \\
LRIS-14 &       09 10 39.6 & 54 21 48 & $0.0000^{+0.0000}_{-0.0000}$ & 0 &     21.74 & $<0.01$ & 20.93 & $<0.01$ & 20.08 & $<0.01$ & 19.66 & $<0.01$ & 19.03 & 0.02 \\
LRIS-33 &       09 10 46.5 & 54 21 41 & $1.1000^{+0.0001}_{-0.0001}$ & 0 &     24.66 & 0.04 & 23.70 & 0.01 & 22.94 & 0.01 & 22.04 & 0.01 & 20.68 & 0.06 \\
LRIS-40 &       09 10 41.4 & 54 20 33 & $0.6299^{+0.0002}_{-0.0002}$ & 0 &     23.27 & 0.01 & 22.45 & 0.01 & 22.02 & 0.01 & 21.78 & 0.01 & 20.84 & 0.06 \\
LRIS-42 &       09 10 48.0 & 54 22 21 & $0.0000^{+0.0000}_{-0.0000}$ & 0 &     22.73 & 0.01 & 22.53 & 0.01 & 22.33 & 0.01 & 21.83 & 0.01 & 20.44 & 0.05 \\
LRIS-47 &       09 10 45.9 & 54 22 20 & $1.1047^{+0.0006}_{-0.0024}$ & 0 &     26.04 & 0.13 & 24.18 & 0.02 & 23.65 & 0.02 & 22.50 & 0.02 & 20.69 & 0.06 \\
LRIS-49 &       09 10 50.0 & 54 22 18 & $0.2205^{+0.0001}_{-0.0002}$ & 0 &     22.23 & $<0.01$ & 22.04 & $<0.01$ & 21.94 & $<0.01$ & 21.98 & 0.01 & 21.72 & 0.12 \\
LRIS-60 &       09 10 41.1 & 54 22 19 & $0.4352^{+0.0001}_{-0.0001}$ & 0 &     22.88 & 0.01 & 22.45 & 0.01 & 22.30 & 0.01 & 22.15 & 0.01 & 21.43 & 0.10 \\
\hline
\end{tabular}
\end{tiny}
\end{table*}

%-------------------------------------------------------
\subsection{X-ray Observations}

The field has been serendipitously covered by XMM-Newton observations
of XY UMA on March 28 2005. The observational ID is 02009601.
Medium filter has been used for pn, MOS1 and MOS2 detectors. with the
corresponding net time after removal of flared time intervals  of 48.5,
56.9 and 54.3 ksec. PN detector clocking mode is FullFrame. Using the
approach developed in Finoguenov et al. (2007 and in prep.) we have
searched for the extended X-ray sources in the image. In detail, we  
remove read-out artifacts, detector background and sky background from
each detector and perform a sophisticated point source flux removal from
the scales of interest to extended source search using wavelets.
We note that the XMM data cover a wider area than the Chandra data
presented in \citet{stanford02}.
Sixteen extended sources have been found, of which one is identified as
the RDCSJ0910 cluster.
Out of 4 galaxy clumps reported in this paper,
X-ray emissions are detected from three of them and
a filament between the two of them.
Further discussions of these detections are presented in Sect. 4.

%-------------------------------------------------------
\section{Catalog Construction and Photometric Redshifts}
\subsection{Catalog Construction}
We have two K-band images, namely, from WFCAM and from MOIRCS.
Since the field coverage, depth, and seeing sizes are different,
we make two photometric catalogs;
$VRizK$ (Suprime-Cam+WFCAM) and $VRizK_s$ (Suprime-Cam+MOIRCS).
We refer to them as the WFCAM catalog and the MOIRCS catalog, respectively.

\subsubsection*{WFCAM catalog}
The seeing size on the WFCAM image is 1.1\arcsec.
We smooth all the Suprime-Cam images to this seeing size
(note that the original seeing of the optical images is 0.8\arcsec). 
We detect galaxies in the $z$-band since this is deeper than
the WFCAM $K$-band image for red galaxies at $z=1.1$.
We then apply a magnitude cut of $z<25.2$ ($5\sigma$) to the catalog.

\subsubsection*{MOIRCS catalog}
The MOIRCS seeing varies from one field to another, and  we smooth
the images to a common 0.8\arcsec\ seeing size. 
The Suprime-Cam images are not smoothed
because all of them have a seeing size of 0.8\arcsec.
The MOIRCS images are deeper than the WFCAM image and
we detect galaxies in the MOIRCS $K_s$-band.
The exposure times of the $K_s$ band vary from field to field
(see Table \ref{tab:data}).
We further subdivide the catalog into two: MOIRCS-shallow
($K_s<22.9$ for all the fields) and MOIRCS-deep
($K_s<23.3$ and F21 and F23 are not used) to fully exploit the data.

In both catalogs, we use {\sc SExtractor} for the object detection \citep{bertin96}.
We use MAG\_AUTO for total magnitudes and 2\arcsec aperture magnitudes for colors.
The photometric errors are derived from the standard deviation
of sky fluxes in 2\arcsec apertures.  The Poisson noise of each object's flux
is then added in quadrature.
Stars are removed on the basis of their compactness and colors.

As we will show later, we perform statistical subtraction of fore-/background galaxies.
We use the data from the Subaru Deep Field (SDF) for this job.
The data were kindly provided by the SDF team.
We apply the same magnitude cuts to the SDF catalog as the WFCAM and
MOIRCS catalogs to make fair samples of the control field.

%-------------------------
\subsection{Photometric Redshifts}

We apply the photometric redshift (photo-$z$) technique to largely eliminate
fore-/background contamination.
A custom designed photo-$z$ code is developed for this purpose.
The code follows the standard procedure.  Each object is fitted with templates
of galaxy spectral energy distributions and the redshift of the best-fitting
template is used as photo-$z$.

A library of templates is made with \citet{bruzual03} models.
We assume the $\tau$ model\footnote{
A star formation rate is proportional to $\exp(-t/\tau)$.
}
to describe the star formation histories of galaxies.
The parameter $\tau$ ranges from 0 (instantaneous burst) to
$\infty$ (constant star formation rate).
The dust extinction is taken into account except for $\tau=0$ models
assuming the \citet{charlot00} extinction model.
The optical depth in the $V$-band is allowed to vary between 0 and 5.
We assume the \citet{chabrier03} initial mass function and solar and sub-solar
metallicities ($Z=0.02$ and 0.008) for the models.
Although the intergalactic extinction is not particularly important for
the purposes of the paper, we implement the extinction following
\citet{furusawa00} with the extinction formula by \citet{madau95}.
Each model is then convolved with the response functions of the filter,
CCD and atmosphere (we assume airmass $=1$) and a library of synthesized
magnitudes are generated.
Each observed object is then fitted with all the templates and
the best-fitting model is searched for based on the $\chi^2$ statistics.

We add 0.03 mag errors in quadrature to the photometric
errors in all the bands to account for systematic zero point errors.
There is often a mismatch between observations and model templates
and we examine a possible mismatch in our data by offsetting observed data points
and checking the resultant photo-$z$.
We find that the magnitude shifts of
$\Delta V=+0.1$ and $\Delta K=+0.3$ mag minimize the catastrophic failures.
We apply these shifts for the photo-$z$  purpose only
(i.e., magnitudes in the following analyses are not shifted).
These procedures are applied in both WFCAM and MOIRCS catalogs.

\begin{figure}
\centering
\includegraphics[width=8cm]{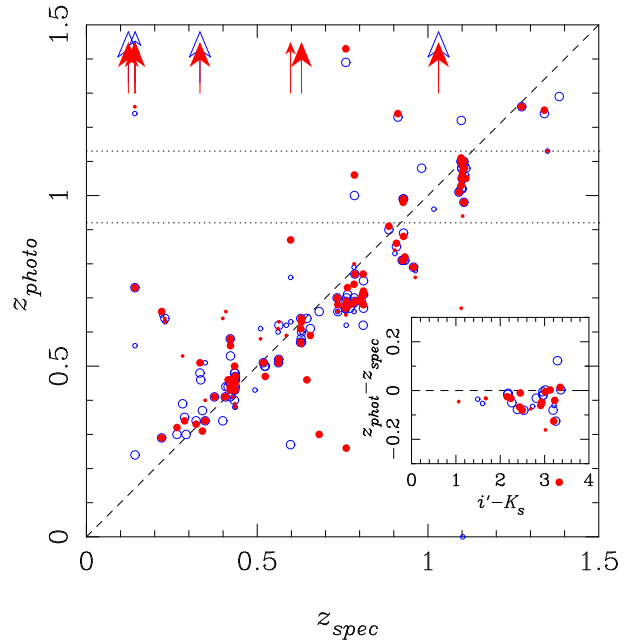}
\caption{
Photometric redshift plotted against spectroscopic redshift.
The open and filled points show photo-$z$ based on
WFCAM and MOIRCS-shallow catalogs, respectively.
As discussed in Sect. 3.1, we make two different photometric
catalogs for the two different sets of near-IR imaging data.
The large/small symbols show galaxies brighter/fainter
than $K=21$.
The open and filled arrows show data points that lie
above the plotted range.
The dashed line shows the $z_{phot}=z_{spec}$ relation.
The horizontal dotted lines show our photo-$z$ selection range
to extract galaxies at $z=1.1$.
The inset shows the color dependence of our photo-$z$
at $1.0<z<1.2$.
}
\label{fig:photoz}
\end{figure}

Fig. \ref{fig:photoz} shows the accuracy of our photo-$z$ estimates.
Strongly blended objects are removed from the plot. 
The data points show a large scatter at $z<0.5$.
This is because we do not have the $U$ and $B$ bands,
which are crucial for low-$z$ objects.
There is no contamination from $z<0.5$ galaxies to $z\sim1.1$.
At $z>0.5$, we tend to underestimate the true redshifts.
But, this not an issue for our purpose of extracting galaxies at $z=1.1$ either
because the systematic offset in the photometric redshifts is known
and we can adjust the photo-$z$ selection range accordingly.

The chosen photo-$z$ range to extract $z=1.1$ galaxies is
a trade off between completeness and contamination.
A narrow range will give us a low contamination rate
with low completeness, and a wide range will give us the opposite.
In this paper, we prefer to be as complete as possible while
maintaining the contamination minimal.
We extract galaxies at $0.92\leq z_{phot}\leq1.12$ as indicated by
the horizontal dotted lines in Fig. \ref{fig:photoz}.
The fraction of $z=1.1$ galaxies missed from this photo-$z$ range
is 11\% and 5\% for the WFCAM and MOIRCS catalogs, respectively.
The fraction of contaminant galaxies (i.e., galaxies outside the
$z=1.1$ structure) within the photo-$z$ range is
26\% and 18\%, respectively.
We statistically subtract this remaining contamination
as detailed below.

We note that the accuracy of photo-$z$ is dependent on magnitudes
and colors of galaxies.
Photo-$z$ becomes less accurate at fainter magnitudes
due to increased photometric errors.
Also, photo-$z$ can be less accurate for blue galaxies \citep{tanaka05,tanaka06}.
We quantify this magnitude and color dependence with the MOIRCS-shallow catalog
(the WFCAM catalog follows a similar trend to the MOIRCS-shallow one).
A fraction of $z=1.1$ galaxies missed from the photo-$z$ range
is 0\% and 17\% for $K_s<21$ and $K_s>21$ galaxies, respectively.
The fraction of contaminant galaxies is 19\% and 17\% for bright
and faint galaxies, respectively.
%Our photo-$z$ indeed appear to become less accurate at faint magnitudes.
We do not observe strong color dependence as shown in the inset,
although we have too few blue galaxies to fully
address the color dependence.
As shown later, the statistical field subtraction without using photo-$z$
gives consistent results to those obtained using photo-$z$.
Therefore, our results presented below are not strongly biased
by photometric redshifts.

%-------------------------------

%-------------------------------------------------------
\section{Large-Scale Structure at $z=1.1$}

\begin{figure}
\centering
\includegraphics[width=8cm]{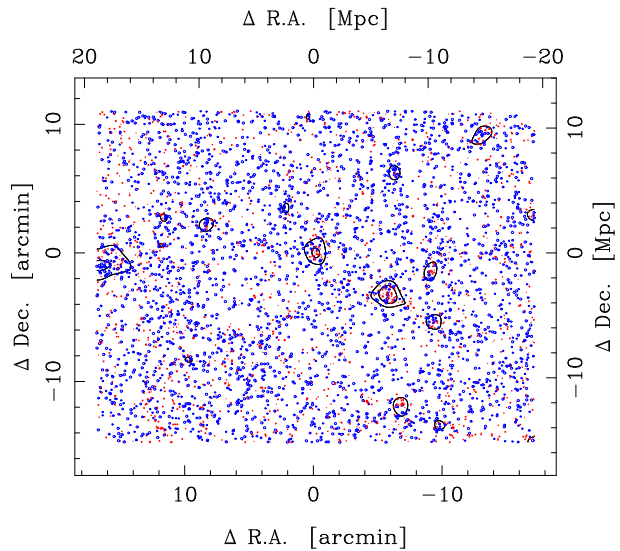}
\caption{
Distribution of the photo-$z$ selected galaxies in the RDCSJ0910 field
($0.92\leq z_{phot}\leq 1.12$).
This is based on the WFCAM catalog.
The galaxies are sorted into red and blue by their $R-z$ color and shown as 
filled and open points, respectively.
The contours show galaxy densities at 2.5, 5, and 10 $\sigma$ levels
estimated by smoothing the galaxy distribution with a 0.5 Mpc Gaussian kernel. 
The RDCSJ0910 cluster lies at ($\Delta$R.A., $\Delta$Dec.)=(0\arcmin,0\arcmin).
The top and right ticks show comoving scales.
}
\label{fig:lss_wfcam}
\end{figure}

We extract galaxies near the cluster redshift and present a
potential large-scale structure around the RDCSJ0910 cluster in Fig. \ref{fig:lss_wfcam}.
The RDCSJ0910 cluster lies at ($\Delta$R.A.,$\Delta$Dec.)=(0\arcmin,0\arcmin).
We discover a cluster candidate at ($\Delta$R.A.,$\Delta$Dec.)=($-6$\arcmin,$-3$\arcmin),
and two more clumps can be seen at NW and SW of this cluster.
These newly found systems show clear concentrations of red galaxies,
which suggest that they are bound systems.
These clumps are closely clustered together,
making it an interesting field at $z=1.1$.

\begin{figure}[hbt]
\centering
\includegraphics[width=8cm]{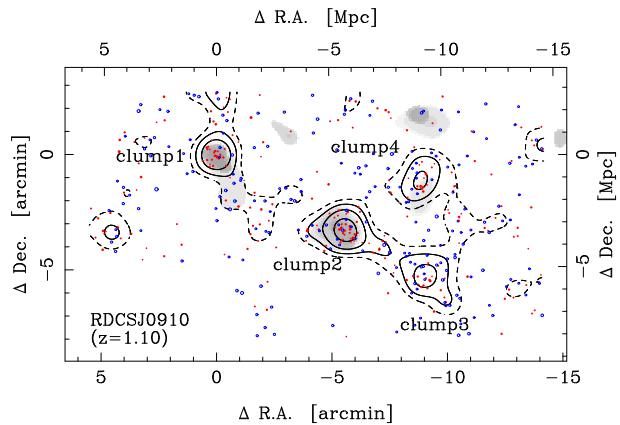}
\caption{
Same as Fig \ref{fig:lss_wfcam} but with the MOIRCS-shallow catalog.
The galaxies are sorted into blue and red by their $i-K_s$ color
following the definition of red galaxies in Sect. 5
and are plotted as the filled and open points.
The dashed contours show $1\sigma$ over-density regions.
The shades show X-ray emissions detected by XMM.
}
\label{fig:lss_moircs}
\end{figure}

\begin{figure*}[tb]
\centering
\includegraphics[width=16cm,height=10cm]{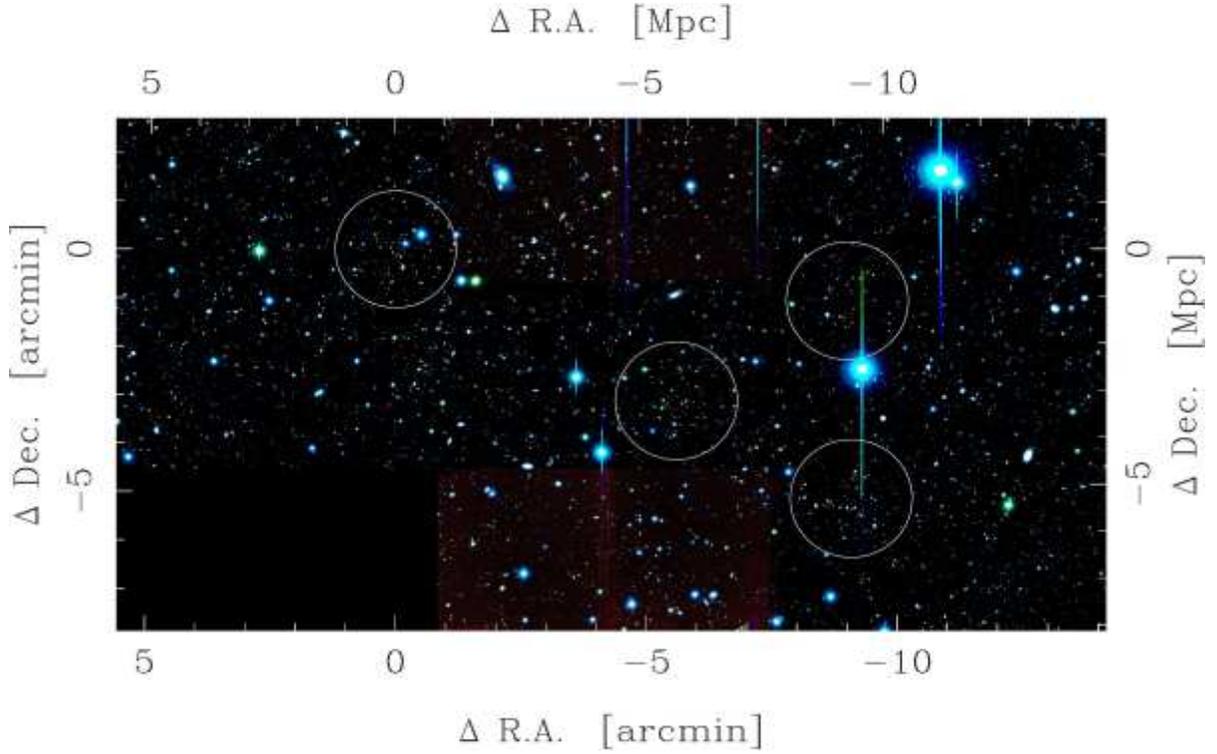}
\caption{
$RzK_s$ pseudo-color image of the RDCSJ0910 field.
The circles show the four clumps.
The top and right ticks show comoving scales.
}
\label{fig:lss_color}
\end{figure*}

Fig. \ref{fig:lss_moircs} shows the same structure in the MOIRCS field.
Refer to Fig. \ref{fig:data} for the overlap of the WFCAM and MOIRCS fields. 
Fig. \ref{fig:lss_color} presents a pseudo-color picture of the field.
The MOIRCS images cover all the newly found systems.
We dub the clumps number 1 to 4 as shown in the figure.
Since the photo-$z$ with the MOIRCS catalog is more accurate
than the WFCAM catalog for faint galaxies (faint galaxies are not detected
in the WFCAM image and hence do not have $K$-band photometry)
and galaxies are detected in the $K_s$ band (which suppresses foreground contamination),
the structure is more clearly seen.

The photo-$z$ selected galaxies seem to show filaments in between the clumps
(see the $1\sigma$ density contours in Fig. \ref{fig:lss_moircs}).
The filament around clump 1 extends in the N-S direction and it bends
south towards clump 2.
Possible filaments can be also seen between clump 2, 3, and 4.
There seems to be a clump to the southeast of clump 1.
However, this is not seen in the WFCAM map or X-ray map
(Figs. \ref{fig:lss_wfcam} and \ref{fig:lss_moircs})
and it is close to the field edge.
We do not discuss this clump further in the following.

An effective way to see if the systems are physically bound or not is to
look for extended X-ray emissions.
We show in Fig. \ref{fig:lss_moircs} X-ray emissions as shades.
The X-ray properties of the clumps are summarized in Table \ref{tab:xray}.
The RDCSJ0910 cluster is clearly detected in X-ray.
The newly found clump 2 also shows an extended emission and
has a similar mass to the RDCSJ0910 cluster mass,
$\rm M_{500}\sim9\times10^{13}M_\odot$, assuming that it lies at $z=1.1$.
Clump 4 is also detected and has a smaller mass of
$M_{500}\sim5\times10^{13}\rm M_\odot$.
The remaining one, clump 3, is not detected at a $3\sigma$ level.
Therefore, at least two of the three new systems are physically bound.

We are not yet sure if the newly discovered clumps lie at the same redshift
and form a single structure due to the limited accuracy of photometric redshifts.
We have carried out spectroscopic campaigns of this field, but
so far we have focused on the main RDCSJ0910 cluster (clump1) and we have not yet
obtained spectra of galaxies in the newly found clumps.
Although this is likely a real structure given that there seems to be
filaments in between the clumps, we defer a firm conclusion on this point
to future spectroscopic observations.

In the rest of the section, we discuss spectroscopically confirmed members
in the main RDCSJ0910 cluster.
Fig. \ref{fig:spec_distrib} shows the distribution of
spectroscopic galaxies.  We have carried out intensive spectroscopic
follow-up observations around the RDCSJ0910 cluster.
The LRIS observations are described in \citet{mei06a}.
We have obtained 20 cluster members, out of which 10
lie within $r_{200}$ from X-ray.
The members are indicated by the stars in the plot.
The N-S extension around the cluster is also seen in the distribution of
spectroscopically confirmed members.
It is interesting to note that the filament towards the SW is also seen in the X-ray data 
(Fig. \ref{fig:lss_moircs}).

Using galaxies at $1.07<z_{spec}<1.13$, the redshift and velocity dispersion
of the RDCSJ0910 cluster is $z=1.1005\pm0.0016$ and $716 \pm 141\rm\ km\ s^{-1}$, respectively.
We use the biweight estimator for these estimates
and the errors are jackknife errors \citep{beers90}.
We also quote some relevant numbers here: $r_{200}=0.95 \pm 0.19$ Mpc,
and $\rm M_{200}=3.4^{+2.4}_{-2.1}\times10^{14}\rm M_\odot$ \citep{carlberg97}.
These numbers appear larger than those estimated with X-ray
($r_{200}=0.59$ Mpc and $\rm M_{200}=1.2\times10^{14} M_\odot$),
suggesting that the cluster may have substructure and/or dynamically young.

To sum up this section, we have discovered the potential clumpy and filamentary
structure around the RDCSJ0910 cluster based on the wide-field multi-band data.
If confirmed to be real, this will be among the few supercluster scale structures
found so far in the $z>1$ Universe.

%-------------------------------

%-------------------------------
\begin{table*}
\caption{X-ray properties of the four clumps (see Fig. \ref{fig:lss_moircs}
for the definition of the clumps).  Clump 3 is not detected in the X-ray data,
and the numbers quoted below are $3\sigma$ limits.
The coordinates correspond to the peak of the X-ray emission. 
The $L_X$ luminosity is based on the extrapolation of the detected flux to
$R_{500}$, described in \citet{finoguenov07}. The total mass is computed
using the scaling relations ($L_X-M$) with the weak lensing calibration for
mass from COSMOS, achieved at a similar redshift ($z\sim0.9$) and
luminosity range (Leauthaud et al. in prep.).
The newly discovered clumps are assumed to lie at the same redshift as the RDCSJ0910
cluster (i.e., clump 1).
}
\label{tab:xray}
\centering
\begin{tabular}{l l l l l }
\hline\hline
clump & R.A.  & Dec.  &  $L_X\rm\ (10^{43}\ ergs\ s^{-1})$ &  $\rm M_{500}\ (10^{13}M_\odot)$ \\
\hline
1 & $09^h\ 10^m\ 45^s$ & $54^\circ\ 22\arcmin\ 08\arcsec$ & $7.3\pm0.9$ & $8.9\pm1.0$ \\
2 & $09^h\ 10^m\ 09^s$ & $54^\circ\ 18\arcmin\ 56\arcsec$ & $7.5\pm1.1$ & $9.0\pm1.1$ \\
3 & $09^h\ 09^m\ 44^s$ & $54^\circ\ 20\arcmin\ 22\arcsec$ & $<3.4\pm1.1$ & $<4.8\pm1.2$ \\
4 & $09^h\ 09^m\ 45^s$ & $54^\circ\ 16\arcmin\ 50\arcsec$ & $3.8\pm1.3$ & $5.2\pm1.4$ \\
\hline
\end{tabular}
\end{table*}

%-------------------------------------------------------
\section{Environmental Dependence of Galaxy Colors at $z=1.1$}

We examine the environmental dependence of galaxy colors at
$z\sim1.1$ in this section.
First, we study the 'traditional' color-radius relation.
We have covered the surrounding regions of the newly found clumps,
and we take this opportunity to examine colors of galaxies out to
large radii at this high redshift.
We then focus on color-magnitude diagrams (CMDs) of galaxies in the clumps.

%-------------------------------------------------------
\subsection{Color-Radius Relation}

We extract $z\sim1.1$ galaxies with photo-$z$ in the MOIRCS-shallow catalog.
Since the masses of the newly found clumps differ from one another,
we take $r_{200}$ estimated using X-ray to normalize the cluster-centric radius.
The clump 3 is not detected in X-ray and we take the $3\sigma$ limit
for this clump.

We plot in Fig. \ref{fig:fred_radius} the fraction of red galaxies
as a function of distance from the clump centers in units of $r_{200}$.
Red galaxies are defined as those with $\Delta|i-K_s|<0.5$
from the red sequence (see below).
In this plot, the statistical field subtraction is not performed\footnote{
A significant fraction of galaxies at large radii is expected to be
fore-/background galaxies, and the subtraction of them results in
a large scatter with large errors in the data points.
Thus, we do not subtract them and instead we draw the red fraction
of the control field as the horizontal line.}.
The fraction stays constant at $r/r_{200}\gtrsim1$ and it increases
at smaller radii reaching to 80\% in the cores.
Galaxies in the cores of clusters at $z\sim1.1$ are almost exclusively red.
The fraction sharply changes at $\sim r_{200}$ and
it converges to the field value at this radius.
The fraction at large radii is slightly higher than the control field value.
This may suggest that galaxies in the filaments have a higher red fraction.

We also plot galaxies from the Sloan Digital Sky Survey for a $z=0$ counterpart
\citep{york00}.  The data are taken from the sixth data release \citep{adelman07}.
We retrieve galaxies at $0.020<z<0.035$ from the Main Galaxy Sample \citep{strauss02}.
To mimic the selection bias in the galaxies in the RDCSJ0910 field,
we derive magnitudes in the $i$ and $K_s$ bands redshifted to $z=1.1$ using
{\sc kcorrect} (v4\_1\_4; Blanton et al. 2003).
We fit the Schechter function to the redshifted $K_s$ band luminosity function
and derive $M_{K_s, z=1.1}^*=-21.96$.
We then apply a magnitude cut of $M_{K_s, z=1.1}<M_{K_s, z=1.1}^*+2.4$
to cover the same magnitude range as the $z=1.1$ sample\footnote{
The $K_s$-band characteristic magnitude of $z\sim1.1$ galaxies is $m_{K_s}^*=20.5$
\citep{strazzullo06} and we reach $m_{K_s}^*=22.9$ ($=m_{K_s}^*+2.4$)
with the MOIRCS-shallow catalog.
We apply the same magnitude cut relative to the characteristic magnitude
to the sloan catalog.
}.

Galaxy groups are identified via the friends-of-friends algorithm
\citep{ramella97,diaferio99,merchan02}.
For the line-of-sight and transverse linking lengths, we adopt $500\rm\ km\ s^{-1}$
and 500 kpc, respectively.
This technique is applied to the volume-limited sample and the linking
lengths do not change with redshift.
Systems with more than 5 members are identified.
These are the same parameters we adopted in \citet{tanaka05}.
For each system, the velocity dispersion and $r_{200}$ are estimated with the
gapper method \citep{beers90}.
Low mass systems ($\sigma<500\rm\ km\ s^{-1}$, which roughly corresponds to
$M_{200}<10^{14}\rm M_\odot$) are removed for a fair comparison with the $z=1.1$ systems.

In the local Universe, the fraction of red galaxies
converges to the field value around $\sim1.5 r_{200}$, being consistent with
earlier studies (e.g., \citealt{carlberg97,tanaka04,rines05}).
Compared to this local value, the $z=1.1$ systems converge to the field value
at smaller radii ($\sim r_{200}$).
But, this may be due to contamination of fore-/background galaxies
since the fraction of contaminant galaxies increases in the outskirts.
The red fraction in the field is higher at $z=0$ than at $z=1.1$, reflecting
the decrease in the cosmic star formation rate.
In contrast to this, the red fraction in the cores does not strongly
evolve since $z=1.1$.
The cluster cores at $z=1.1$ are already dominated by
red galaxies.

\begin{figure*}[hbt]
\centering
\includegraphics[width=12cm]{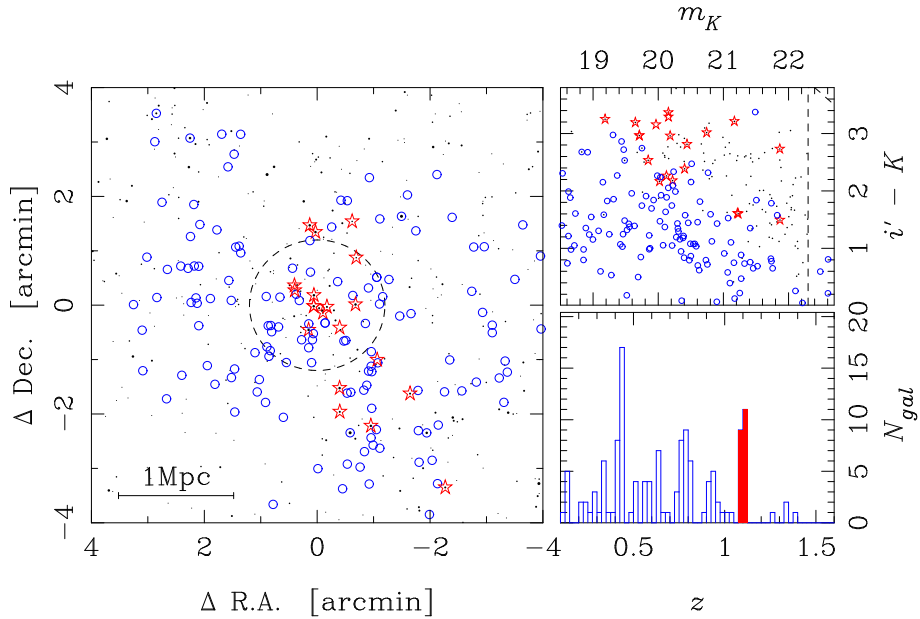}
\caption{
The distribution of galaxies with spectroscopic and photometric redshifts
around the RDCSJ0910 cluster.
{\it Bottom-right:} Redshift distribution of the spectroscopic galaxies.
The filled histogram shows the $z=1.1$ structure.
Galaxies inside/outside the filled histogram are shown as
stars/open circle in the other panels.
{\it Top-right:} $i-K$ color plotted against $K$ magnitude.
The black dots are photo-$z$ selected galaxies.
The dashed lines show the $5\sigma$ magnitude and color.
{\it Left:} Spatial distribution of the spectroscopic galaxies.
The dashed circle shows $r_{200}$ of the cluster estimated from
our X-ray data.  The 1 Mpc physical scale is shown at the bottom left.
}
\label{fig:spec_distrib}
\end{figure*}

Let us be more quantitative about this trend.
We find that the fraction of red galaxies corrected for the field contamination
in the combined clumps 1-4 within $r_{200}$ is $0.72\pm0.03$.
The $z=0$ rich systems have a similar fraction of $0.82\pm0.05$.
The red fraction shows only a small change since $z=1.1$.
If we focus on rich systems (clumps 1 and 2), the red fraction
is $0.80\pm0.04$, which is consistent with the local value.
This suggests that red galaxies have already become a dominant population
in rich clusters at $z=1.1$ and no significant evolution is seen
in the fraction of red galaxies.
In other words, the environmental dependence of galaxy colors is already
fully in place at $z=1.1$.
As discussed in \citet{tanaka06}, the accuracy of photometric redshifts
may depend on galaxy color so we may be missing some blue galaxies.
Our current spectroscopic sample does not allow us to examine
this point in detail because we have few spectroscopic galaxies at $z=1.1$
with blue colors (see Fig. \ref{fig:spec_distrib}).
But, we apply the statistical field subtraction without using
photo-$z$ and find a similar red fraction : $0.77\pm0.08$ for
clump 1+2.
Therefore, this high fraction is not due to possible photo-$z$ biases,
but is likely robust.

The poor systems (i.e., clumps 3 and 4) show a smaller red fraction of $0.59\pm0.05$
(the statistical subtraction gives $0.61\pm0.10$, again being consistent
with that obtained with photo-$z$).
The fraction of red galaxies is $\sim35\%$ higher in rich systems.
Because we use the same data set for these two combined systems,
the relative comparison between clumps 1 and 2 and clumps 3 and 4 is robust.
A dependence of the stellar populations in the probable member galaxies
on cluster richness also appears to be in place already at $z=1.1$.

%The trends we see here need to be confirmed with a larger sample,
%but the implication is that the strong environmental
%dependence of galaxy colors seen at $z < 1$ is already in place, at least qualitatively, at $z = 1.1$.
%We will further discuss this point in Sect. 7.

%-------------------------------

%-------------------------------

%-------------------------------------------------------
\subsection{Color-Magnitude Diagrams}

We now turn our attention to the color-magnitude diagrams of the galaxies.
Since we have four galaxy systems at $z\sim1.1$ with various masses,
it is interesting to examine variations of galaxy colors and magnitudes
in different systems.
In this subsection, we use the MOIRCS-deep catalog.
Note that this catalog covers all the clumps.

\begin{figure}
\centering
\includegraphics[width=8cm]{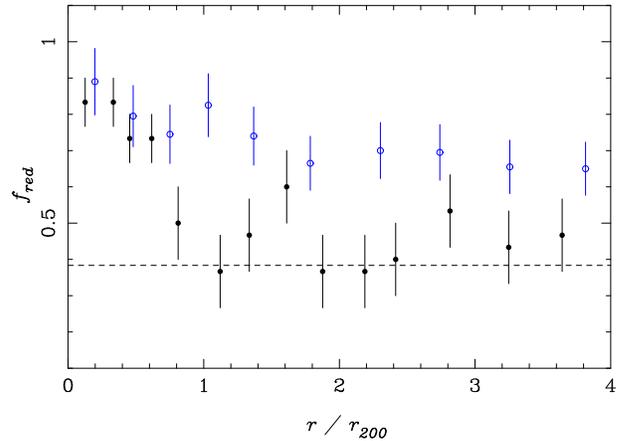}
\caption{
Fraction of red galaxies plotted against cluster-centric radius.
The filled points show galaxies at $z=1.1$ and  each bin contains 50 galaxies.
The error bars show the Poisson errors.
The horizontal dashed line shows the red fraction of the control field sample.
Statistical field subtraction is not performed in this plot.
The open points show $z=0$ galaxies and each bin contains 200 galaxies.
}
\label{fig:fred_radius}
\end{figure}

\begin{figure}
\centering
\includegraphics[width=8cm]{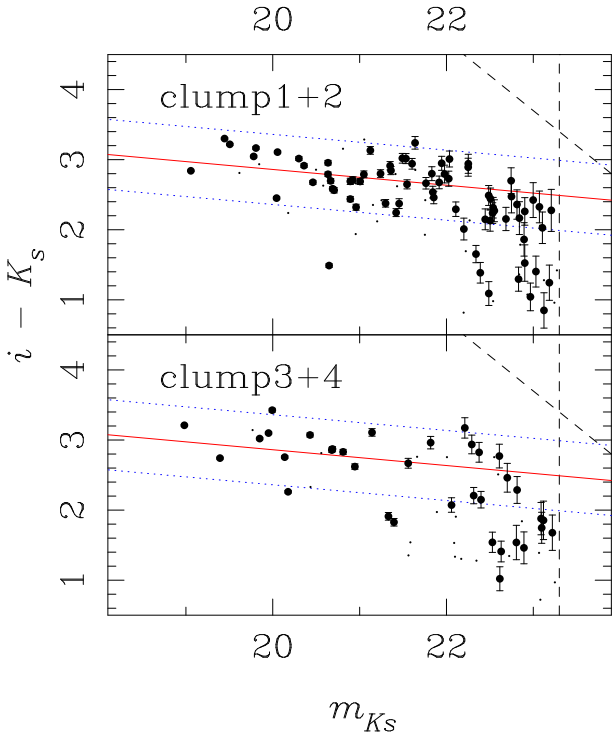}
\caption{
$i-K_s$ color plotted against $K_s$ magnitude.
The dashed line shows the $5\sigma$ limiting magnitudes and colors.
The solid line is the best-fitting red sequence in clump1+2.
The dotted lines show $\Delta(i-K_s)=\pm0.5$ from the red sequence.
The small dots are statistically subtracted galaxies.
}
\label{fig:cmd}
\end{figure}

Fig. \ref{fig:cmd} shows the CMDs of the clumps 1-4.
We extract galaxies within $r_{200}$ from the average center of
the few brightest galaxies in each clump.
$r_{200}$ for each clump is estimated from the X-ray data.
For clump3, we use the $3\sigma$ upper limit on $r_{200}$ since
this is not detected at a significant level.
$r_{200}$ effectively covers the apparent extent of each clump
as shown in Fig. \ref{fig:fred_radius}.
We make a combined clump of the clumps 1 and 2, both of which have
similar masses (see Table \ref{tab:xray}). 
We do the same for clump 3 and 4.

A statistical field subtraction is performed in this plot
following the procedure detailed in \citet{tanaka05}.
In short, the expected number of contaminant galaxies
is estimated based on the average surface density of the control galaxies.
The distribution of the control galaxies in the CMD is
used as a probability map of contamination and the galaxies in
the clumps are statistically subtracted according to their
field probabilities.
Fig. \ref{fig:cmd} is one such realization of the statistical subtraction.

A clear sequence of red galaxies is observed in clumps 1 and 2.
We apply the bi-weight fit to the sequence at $K_s<22$ and obtain
$i-K_s=-0.112^{+0.051}_{-0.049}\times(K_s-20)+2.86^{+0.06}_{-0.06}$.
The errors are estimated by bootstrapping the input catalog.
We take this as the best-fitting red sequence and plot it in Fig. \ref{fig:cmd}.
The slope is slightly steeper than our earlier measurement
for the RDCSJ1252 cluster at $z=1.24$ (slope of the $i-K_s$ color is
$-0.058^{+0.026}_{-0.032}$; \citealt{tanaka07a}),
but the slope observed here fits the RDCSJ1252 red sequence reasonably well.
Galaxies at $K_s>22$ in clumps 1 and 2 tend to be on the blue side
of the sequence.
One can fit a steeper slope using all the red galaxies down to
the magnitude limit:  slope $=-0.195^{+0.03}_{-0.03}$.
But, this does not fit the RDCSJ1252 red sequence, suggesting that
those faint red galaxies in the clump1+2 may not have become fully red yet.

The red sequence in clumps 3 and 4 at $K_s<22$ shows similar slope
and offset to the clump1+2 red sequence:
$i-K_s=-0.099^{+0.072}_{-0.072}\times(K_s-20)+2.92^{+0.08}_{-0.08}$.
The location of the red sequence on a CMD indicates the redshift of a system.
Higher redshift clusters form an apparently redder sequence.
The similarity of the locations of the sequences in these clumps
suggests that they lie at similar redshifts.

To be quantitative about the red sequence,
we measure the color scatter around the red sequence
and show them in Table \ref{tab:color_scatter}.
The scatter is measured as follows.  The input catalog is bootstrapped and
the statistical field subtraction is performed.
Then, the surviving galaxies within $|i-Ks|<0.75$ of the red sequence are extracted.
This color range is wide enough to include all red galaxies, while it is
small enough to remove most of the blue galaxies.
We apply $2\sigma$ clipping once to exclude outliers and
measure the intrinsic scatter assuming
$\sigma_{measured}^2=\sigma_{intrinsic}^2+\sigma_{photometric\ error}^2$.
This procedure is repeated 100,000 times and we take the median of the
distribution of the color scatter as the estimate and
a 68\% interval as its error.

The scatter does not significantly increase down to the faintest magnitude bin
in clumps 1 and 2, suggesting that the red sequence is already formed.
\citet{mei06a} observed the tight red sequence in the $i-z$ color
down to $z=24$, which corresponds to $K_s\sim22$ for red galaxies at $z=1.1$,
in the RDCSJ0910 cluster (i.e., clump 1).
Our result appears consistent with their work, but we do not make
any morphological selection, nor spectroscopic selection in this work.
Dusty spirals and fore-/background contamination can increase
the apparent color scatter here.
Also, the $i-K_s$ color is more sensitive to age and dust effects
than the $i-z$ color.
Therefore, the comparison between their result and our result here
is not straight forward.
There is a hint of an increased scatter at $K_s>21.3$ in the clump3+4,
although this is not statistically significant.

\begin{figure}
\centering
\includegraphics[width=8cm]{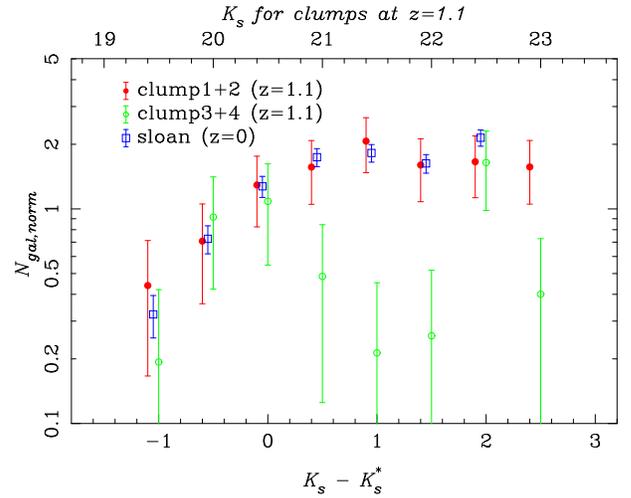}
\caption{
LF of red galaxies.
The filled points, open points, and open squares show the clump1+2, clump3+4,
and the sloan data (i.e., $z=0$) respectively.
The magnitudes are plotted relative to the characteristic magnitudes
at $z=1.1$ for the clumps and $z=0$ for the sloan data.
We note that the sloan photometry is $k$-corrected to probe the same
rest-frame wavelength as the $K_s$ band for $z=1.1$ galaxies.
The LFs are normalized at $-0.5<K_s-K_s^*<0$.
The top ticks show $K_s$ magnitudes at $z=1.1$.
Note that the sloan data do not reach as deep as the MOIRCS-deep
catalog. % and the faintest magnitude bin is not plotted.
The points are slightly shifted horizontally in order to avoid overlapping.
The error bars show the Poisson error.
}
\label{fig:lf}
\end{figure}

To further quantify the red sequence, we plot in  Fig. \ref{fig:lf}
the luminosity function (LF) of red galaxies in the clumps 1+2 and 3+4.
Here the red galaxies are defined as those within the $\Delta|i-K_s|<0.5$
from the best-fitting red sequence
(i.e., those between the dotted lines in Fig. \ref{fig:cmd}).
The LFs are normalized at $K_s=20.0-20.5$ ($\sim M_{K_s}^*$,
\citealt{strazzullo06}) for comparison.
The LF of clump1+2 is a smooth function with a hint of
a slight decrease at the faint end ($K_s\gtrsim22$).
In contrast to this, the number of the red galaxies
in the clump3+4 starts to decrease at a brighter magnitude of $K_s\sim21$
with some scatter at fainter magnitudes.
This suggests that the red sequence is not fully in place
at faint magnitudes in the clump3+4.

We also plot red galaxies at $z=0$ from the Sloan survey as the open squares
in Fig. \ref{fig:lf}.
We recall that we $k$-corrected the sloan photometry to probe the same
rest-frame wavelength as the $K_s$-band for $z=1.1$ galaxies.
The magnitudes are plotted relative to the characteristic magnitudes
at each redshift (the top ticks show apparent $K_s$ band magnitudes
of $z=1.1$ galaxies).
The LF of clump1+2 is consistent with the sloan data, suggesting
that the red sequence is fully built up by $z\sim1$ in rich clusters
at least down to $K_s^*+2$ and massive cluster galaxies are
already fully assembled.
In contrast to this, the clump3+4 shows a clear deficit of
faint red galaxies.  Faint galaxies are not yet assembled
and/or are still blue in poor groups.

Another interesting way of quantifying the red sequence will be
the ratio of luminous red galaxies to faint red galaxies.
We define luminous and faint galaxies as $m_{K_s}<21$ and $21<m_{K_s}<23.3$.
The luminous to faint galaxies ratio is $0.48\pm0.12$,
and $1.06\pm0.47$ for clump1+2 and 3+4, respectively.
Poorer systems tend to lack faint red galaxies.

As a sanity check, we derive LFs of red galaxies using the statistical
field subtraction technique only.  The LFs agree with those obtained
with photo-$z$ within the errors.  The luminous to faint ratio is
$0.51\pm0.11$ and $1.43\pm0.46$ for clump1+2 and 3+4, respectively.
The ratios are again in consistent within the error.
We therefore suggest that our results are not strongly affected
by biases in photometric redshifts.

The faint end of the red sequence is not fully in place in poor groups
at high redshifts ---
this is consistent with our earlier work \citep{tanaka05,tanaka07a}.
The red sequence in the clump3+4 seems to be truncated at $K_s\sim21$.
Interestingly, this truncation magnitude is brighter than that we found
in the RDCSJ1252 cluster at $z=1.24$.
In \citet{tanaka07a}, we discovered four poor groups around the RDCSJ1252 cluster at $z=1.24$.
These groups have $\rm M_{500}\sim4\times10^{14}\rm M_\odot$,
which is similar to the clump4 mass.
But, the red sequence in these groups is sharply truncated at $Ks=22$, which
is fainter by $1$ mag. than that we find here.
This suggests that there is a large variation in evolutionary phases
of galaxies in $z\sim1$ groups even if galaxies live in
groups of similar masses.
We defer discussions on implications of these results for galaxy evolution
to Section 7.

%-------------------------------

%-------------------------------

%-------------------------------

%-------------------------------
\begin{table}
\caption{
The intrinsic $i-K_s$ color scatter around the red sequence
as a function of $K_s$ band magnitude.
}
\label{tab:color_scatter}
\centering
\begin{tabular}{l c c}
\hline\hline
magnitude  & clump1+2  & clump3+4 \\
\hline
$19.3 - 20.3$ & $0.281^{+0.040}_{-0.021}$ & $0.313^{+0.042}_{-0.025}$ \\
$20.3 - 21.3$ & $0.256^{+0.024}_{-0.025}$ & $0.248^{+0.027}_{-0.052}$ \\
$21.3 - 22.3$ & $0.297^{+0.030}_{-0.034}$ & $0.417^{+0.066}_{-0.079}$ \\
$22.3 - 23.3$ & $0.297^{+0.049}_{-0.055}$ & $0.338^{+0.064}_{-0.079}$ \\	
\hline
\end{tabular}
\end{table}

%-------------------------------------------------------
\section{Composite Spectra of Red Galaxies}

\begin{figure}
\centering
\includegraphics[width=8cm]{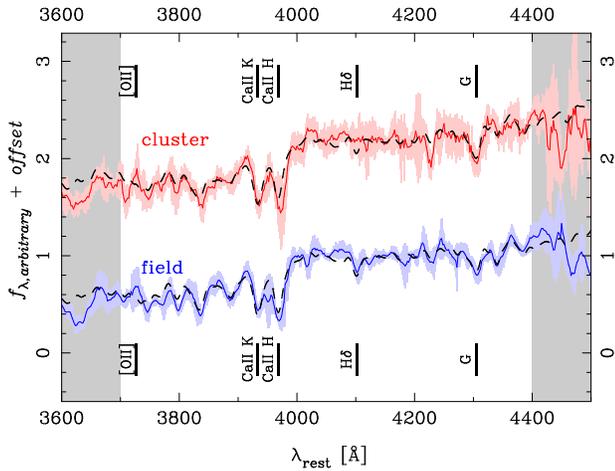}
\caption{
Composite spectra of red galaxies within $r_{200}$ from
the clump1 center (cluster) and outside  $r_{200}$ (field).
The shades associated with the spectra show $1\sigma$ errors.
The dashed lines show the best-fitting passive evolution model.
The shaded regions on the right and left of the plot are not
used in the fit since these regions are strongly affected by
telluric absorptions and it is hard to calibrate fluxes well
at long wavelengths ($\lambda_{obs}>9000\rm\AA$).
}
\label{fig:spec}
\end{figure}

In this section, we study the spectroscopic properties of galaxies in the clumps.
We present the composite spectrum of the RDCSJ0910 cluster (i.e., clump1)
red galaxies in Fig. \ref{fig:spec}.
The red galaxies are defined as those within $r_{200}$ (derived from the X-ray data) 
with colors of $|i-K_s|<0.5$ from the red sequence.
We exclude one AGN from the X-ray data presented in \citet{stanford02}
and \citet{mei06a} from the following analysis.
We smooth the spectra to a common instrumental resolution of $R\sim350$.
Each spectrum is normalized to unity at $4000-4200\rm\AA$ and
8 spectra are combined by taking a $2\sigma$-clipped mean.
For comparison, we also make the composite spectrum of red galaxies
outside $r_{200}$ using 5 galaxies.
In what follows, we refer to the two composite spectra as cluster and field spectra
for simplicity.
It should be noted that some of the red galaxies outside $r_{200}$
may be bound to the cluster and they may not be representative of
the field galaxies.

Both spectra show a strong $4000\rm\AA$ break with prominent
Ca{\sc ii}HK lines, suggesting that galaxies are dominated by evolved stars.
Their red colors are not due to strong dust extinctions.
Although {\sc [oii]} emissions are seen in a few individual spectrum,
the composite spectra show
no strong {\sc [oii]} emissions and apparently
red galaxies in and around clusters at $z=1.1$ are not actively forming stars.
Most at $z=1.1$ have already suppressed their star formation.

This fact motivates us to fit a passive evolution model to the spectra
and derive an average age of the red galaxies in two different environments.
We use the \citet{bruzual03} model assuming passive evolution,
solar metallicity, Chabrier IMF, and no dust extinction.
The best-fitting model spectra is shown along with the observed composite
spectra in Fig. \ref{fig:spec}.
The model fits the overall continuum shape and some of the most prominent
absorption lines such as G-band well.
However, some differences between models and observations can be seen.
For example, the Ca{\sc ii}HK absorptions are not fit very well by the models.
Most of such differences will probably be due to strong sky lines in this wavelength region.
We stack only 8 and 5 galaxies, and the effects of OH emissions
cannot be completely removed.

The input spectra are bootstrapped and the models are fit for
each composite spectrum.  Then, the median and a 68\% interval of
the best-fitting age distribution are taken as an estimate and its error.
The best-fitting age for the cluster red galaxies is $3.5^{+1.0}_{-0.5}$ Gyr,
which suggests that the bulk of stars in the red galaxies formed
at $z_f=3.3^{+2.9}_{-0.6}$.
This seems to be consistent with earlier studies of high redshift clusters
(e.g., \citealt{blakeslee03,lidman04,holden04,mei06a,lidman08,mei08}).
In contrast to the cluster red galaxies, the field red galaxies show
a younger age of $2.0^{+1.0}_{-1.0}$ Gyr, or $z_f=1.9^{+0.8}_{-0.5}$.
The field red galaxies are younger than the cluster red galaxies by 1.3 Gyr,
although this is a statistically marginal difference.
This age difference is roughly consistent with earlier studies (e.g., \citealt{thomas05}).
We note that the age-sensitive H$\delta$ absorption is stronger
in the field red galaxies, which also suggests their younger age.

We note that the broad-band properties of these red galaxies
in-/outside $r_{200}$ are slightly different.
The mean and dispersion of the $K_s$-band magnitudes and
$i-K_s$ colors of these galaxies are $K_s=20.4\pm0.7$ and $i-K_s=2.94\pm0.25$
and $K_s=19.9\pm0.5$ and $i-K_s=3.07\pm0.29$ for cluster and field galaxies,
respectively.
The Mann-Whitney test shows that the probability that
the magnitudes and colors are drawn from the same population is 10\%.
The colors of cluster and field red galaxies are consistent within
the dispersion and the field red galaxies have slightly redder
average color.
Thus, the younger age of the field red galaxies is not due to
color selection biases, but likely a real trend.

It would be interesting to see if there is any difference in
stellar populations of galaxies in between poor groups and rich clusters.
But, we do not yet have spectra of galaxies in the newly discovered clumps,
and we defer discussions on spectral differences in different environments
to a future paper.

%-------------------------------

%-------------------------------------------------------
\section{Discussions}

\subsection{Comparisons with Previous Work}

One of the primary findings of this work is that the environmental dependence
of galaxy colors seems to be already in place in the $z=1.1$ Universe.
Rich clusters are already dominated by red galaxies and the fraction
of red galaxies within $r_{200}$ at $z=1.1$ is similar to the fraction at $z=0$,
while the red fraction in the field significantly increases down to $z=0$.

Our results might appear inconsistent with earlier work.
For example, 
\citet{cucciati06} reported that the color-density relation weakens at higher redshift.
However, they examined the dependence of colors on large-scale environments
(e.g., they defined environments with a 5Mpc Gaussian aperture), 
and as remarked by them, their results cannot be extrapolated down to cluster scales.

\citet{cooper07} also showed that the color-density relation weakens at higher redshifts
based on the DEEP2 data (see also \citealt{gerke07} for a similar result).
There are two fundamental differences between our sample and theirs.
Firstly, their galaxies are selected in the rest-frame $B$-band,
while we use the $K_s$ band, which corresponds to rest-frame $1\mu m$.
The rest-frame $B$ band detection can be affected by on-going
star formation activity.
They tend to trace blue galaxies and their density estimates are
biased by the density of blue galaxies.
The $K_s$-band selection will probably give a higher
red fraction in high density regions.
Secondly, they did not probe rich clusters, while we do.
%The red fraction is a function of cluster richness as we have shown
%in the paper, and it seems that moderate to high mass
%clusters are not in their sample.
The range of environments explored in this paper is different
from that explored by DEEP2.
Therefore, we cannot make a fair comparison due to these differences
and their results are not in apparent conflict with ours.

\citet{elbaz07} showsed that there is a moderate color-density
relation at $z\sim1$, but their red fraction in the highest
density retion is not as high as we found in this paper.
This is because they did not probe rich clusters.
Recently, \citet{cowie08} found no strong dependence of the red fraction
on environment based on a $K$-band selected catalog.
But, they did not probe rich clusters either.

%---------------------------------------------------
\subsection{Implications for Galaxy Evolution}

We have discovered three new clumps of galaxies near the RDCSJ0910 cluster.
One of which has a similar mass to the RDCSJ0910 cluster mass, while
the remaining two are lower mass systems.
The rich systems at $z=1.1$ seem to be fully evolved systems.
They are dominated by red galaxies and the LF of red galaxies is
nearly flat at faint magnitudes
and is actually consistent with the local LF from the Sloan survey.
This is also consistent with
earlier findings that stellar mass functions of galaxies in rich clusters
do not strongly evolve since $z\sim1$ (e.g., \citealt{strazzullo06}).
On the other hand, the poor groups at $z=1.1$ exhibit a deficit
of faint red galaxies.
The red sequence is not yet fully in place at faint magnitudes
in these systems.
Also, the fraction of red galaxies is lower than clusters.
It seems galaxies in lower density regions evolve more strongly at $z<1$.

We suggested in our earlier papers that galaxy-galaxy interactions
in low density environments such as groups and field may be
the primary physical driver of the environmental dependence \citep{tanaka06,tanaka07b}.
Our results in this paper seem to support this idea.
The fraction of red galaxies increases in poor groups at lower redshifts,
while that in rich clusters does not significantly increase.
It is the group environment where environmental effects act on galaxies.
Progenitors of rich clusters at $z=1.1$ should be poor groups
at higher redshifts and the cluster galaxies have already
changed their properties by the time we observe them at $z=1.1$.
The most likely physical process will be galaxy-galaxy interactions
because they should happen most frequently in groups and
cluster-specific mechanisms such as ram-pressure stripping
are unlikely to play a major role there.
Galaxy-galaxy interactions may trigger intense starbursts,
which may consume gas in galaxies.  The subsequent star formation
activities are suppressed due to lack of gas in galaxies.
However, it will be essential to perform detailed spectral analyses of
the galaxies in the poor groups at $z=1.1$ to prove the scenario.

We now turn our attention to another interesting finding of this work.
The rich systems exhibit the red sequence down to faint magnitudes,
while the poor systems lack faint red galaxies.
This is fully consistent with the picture of the build-up of
the red sequence drawn in our earlier papers and those by other authors
(\citealt{tanaka05,tanaka07a,koyama07,gilbank08}, but see also \citealt{delucia07}).
The red sequence is first built-up at bright magnitudes and
extends to fainter magnitudes, and this build-up is 'delayed'
in poor systems.

But, there seems to be a large variation in evolutionary
phases of galaxies at this redshift.  We have observed that the red sequence
in poor systems around RDCSJ0910 is truncated around $K_s=21$.
In \citet{tanaka07a}, we reported that the red sequence is sharply truncated
at $K_s=22$ in poor systems around the RDCSJ1252 cluster at $z=1.24$.
Although these systems have similar masses, the truncation magnitude appears
to differ by $\Delta K_s\sim1$.
It seems that galaxies in the RDCSJ1252 groups are more evolved than
those in the RDCSJ0910 groups, although the RDCSJ1252 lies at higher redshift
(the time difference between $z=1.24$ and $z=1.1$ is $\sim0.5$ Gyr).

We speculate that initial conditions of galaxy formation come into play here.
The evolution of galaxies is determined by both nature and nurture effects.
Our finding that galaxy properties can be quite different at high redshifts
even if galaxies live in groups of similar masses
seems to suggest that the nature effect plays a role in the $z>1$ Universe.
The structure around the RDCSJ1252 cluster at $z=1.24$ is somewhat loose \citep{tanaka07a}.
On the other hand, the structure around the RDCSJ0910 cluster reported
in this paper is more compact and prominent.
However, galaxies seem to be less evolved in the groups around the RDCSJ0910 cluster.
This is somewhat contrary to what one may expect, but this probably means
that we simply need much better statistics.
There are a good number of $z>1$ cluster candidates in large surveys
such as COSMOS, and statistics will be improved in the near future.
Further work on $z>1$ clusters will be a promising way of addressing
the origin of the environmental dependence of galaxy properties.

%-------------------------------------------------------
\section{Conclusions}

We have carried out the intensive multi-band wide-field observations
of the cluster RDCSJ0910 at $z=1.1$ with Suprime-Cam, WFCAM, MOIRCS,
FOCAS, and LRIS. 
The potential large-scale structure is discovered based on
the photometric redshifts.
Two of the three newly discovered systems show the extended X-ray emissions,
suggesting that these are physically bound systems.
There seems to be filaments in between the clumps.
If confirmed to be real, this will be one of the most prominent structure
ever found in the $z>1$ Universe.

We then look into properties of galaxies in the structure.
The fraction of red galaxies in rich clusters has not changed since $z\sim1$,
suggesting that the environmental dependence of galaxy colors is already in place.
Poor groups at $z\sim1$ show a lower red fraction than rich clusters.

We find that the red sequence of galaxies in relatively poor systems
appears to be truncated at $K_s=21$, while that in richer systems
exhibit a clear sequence down to fainter magnitudes.
This confirms our earlier claim that the red sequence grows from
bright magnitudes to faint magnitudes, and the build-up of the red sequence
happens earlier in richer systems.
Therefore, galaxies follow the environment-dependent down-sizing ---
massive galaxies in higher density environments are more evolved
than less massive galaxies in less dense environments.
Interestingly, the red sequence is truncated at brighter magnitudes
than that found in the groups in the RDCSJ1252 field at $z=1.24$.
Although the groups have similar masses, the truncation
magnitude is $\sim1$ mag brighter in the RDCSJ0910 field at $z=1.1$.

Both the initial conditions of galaxy formation and the environmental effects
determine the fate of a galaxy.
It is difficult to disentangle these two effects, but $z>1$ might be
an era when the nature effects can be seen.
Further detailed studies on $z>1$ clusters would be a promising way of 
addressing this fundamental issue of galaxy evolution.

%-------------------------------------------------------
\begin{acknowledgements}
This study is based on data collected at Subaru Telescope, which is operated by
the National Astronomical Observatory of Japan, and on
observations made with the United Kingdom Infrared Telescope, which
is operated by the Joint Astronomy Centre on behalf of the U.K.
Particle Physics and Astronomy Research Council.
The observations with the UKIRT 3.8-m telescope were supported by NAOJ.
We thank K. Motohara and M. Hayashi for providing the WFCAM $K$-band data
and H. Furusawa and I. Tanaka for their help during
the Suprime-Cam and MOIRCS observations, respectively.
We thank the referee for his/her very helpful comments, which improved the paper.
SAS's work was performed under the auspices of the U.S. Department of Energy,
National Nuclear Security Administration by the University of California,
Lawrence Livermore National Laboratory under contract No. W-7405-Eng-48.
This work was financially supported in part by a Grant-in-Aid
for the Scientific Research (No. 15740126, 18684004) by
the Japanese Ministry of Education, Culture, Sports and Science.
AF acknowledges support from BMBF/DLR under
grant 50 OR 0207, MPG and and a partial support from NASA grant
NNX08AD93G, covering his stays at UMBC.
Some of the data presented herein were obtained at the W.M. Keck Observatory,
which is operated as a scientific partnership among the California Institute
of Technology, the University of California and the National Aeronautics and
Space Administration. The Observatory was made possible by the generous
financial support of the W.M. Keck Foundation.
Funding for the SDSS and SDSS-II has been provided by the Alfred
P. Sloan Foundation, the Participating Institutions, the National Science
Foundation, the U.S. Department of Energy, the National Aeronautics and
Space Administration, the Japanese Monbukagakusho, the Max Planck Society,
and the Higher Education Funding Council for England.
The SDSS Web Site is http://www.sdss.org/.
\end{acknowledgements}

% %-------------------------------------------------------
\bibliographystyle{aa}
\bibliography{0910_refs}

\begin{thebibliography}{56}
\expandafter\ifx\csname natexlab\endcsname\relax\def\natexlab#1{#1}\fi

\bibitem[{{Adelman-McCarthy} {et~al.}(2007){Adelman-McCarthy}, {Ag{\"u}eros},
  {Allam}, {Anderson}, {Anderson}, {Annis}, {Bahcall}, {Bailer-Jones},
  {Baldry}, {Barentine}, {Beers}, {Belokurov}, {Berlind}, {Bernardi},
  {Blanton}, {Bochanski}, {Boroski}, {Bramich}, {Brewington}, {Brinchmann},
  {Brinkmann}, {Brunner}, {Budav{\'a}ri}, {Carey}, {Carliles}, {Carr},
  {Castander}, {Connolly}, {Cool}, {Cunha}, {Csabai}, {Dalcanton}, {Doi},
  {Eisenstein}, {Evans}, {Evans}, {Fan}, {Finkbeiner}, {Friedman}, {Frieman},
  {Fukugita}, {Gillespie}, {Gilmore}, {Glazebrook}, {Gray}, {Grebel}, {Gunn},
  {de Haas}, {Hall}, {Harvanek}, {Hawley}, {Hayes}, {Heckman}, {Hendry},
  {Hennessy}, {Hindsley}, {Hirata}, {Hogan}, {Hogg}, {Holtzman}, {Ichikawa},
  {Ichikawa}, {Ivezi{\'c}}, {Jester}, {Johnston}, {Jorgensen}, {Juri{\'c}},
  {Kauffmann}, {Kent}, {Kleinman}, {Knapp}, {Kniazev}, {Kron}, {Krzesinski},
  {Kuropatkin}, {Lamb}, {Lampeitl}, {Lee}, {Leger}, {Lima}, {Lin}, {Long},
  {Loveday}, {Lupton}, {Mandelbaum}, {Margon}, {Mart{\'{\i}}nez-Delgado},
  {Matsubara}, {McGehee}, {McKay}, {Meiksin}, {Munn}, {Nakajima}, {Nash},
  {Neilsen}, {Newberg}, {Nichol}, {Nieto-Santisteban}, {Nitta}, {Oyaizu},
  {Okamura}, {Ostriker}, {Padmanabhan}, {Park}, {Peoples}, {Pier}, {Pope},
  {Pourbaix}, {Quinn}, {Raddick}, {Re Fiorentin}, {Richards}, {Richmond},
  {Rix}, {Rockosi}, {Schlegel}, {Schneider}, {Scranton}, {Seljak}, {Sheldon},
  {Shimasaku}, {Silvestri}, {Smith}, {Smol{\v c}i{\'c}}, {Snedden}, {Stebbins},
  {Stoughton}, {Strauss}, {SubbaRao}, {Suto}, {Szalay}, {Szapudi}, {Szkody},
  {Tegmark}, {Thakar}, {Tremonti}, {Tucker}, {Uomoto}, {Vanden Berk},
  {Vandenberg}, {Vidrih}, {Vogeley}, {Voges}, {Vogt}, {Weinberg}, {West},
  {White}, {Wilhite}, {Yanny}, {Yocum}, {York}, {Zehavi}, {Zibetti}, \&
  {Zucker}}]{adelman07}
{Adelman-McCarthy}, J.~K., {Ag{\"u}eros}, M.~A., {Allam}, S.~S., {et~al.} 2007,
  \apjs, 172, 634

\bibitem[{{Balogh} {et~al.}(2002){Balogh}, {Smail}, {Bower}, {Ziegler},
  {Smith}, {Davies}, {Gaztelu}, {Kneib}, \& {Ebeling}}]{balogh02}
{Balogh}, M.~L., {Smail}, I., {Bower}, R.~G., {et~al.} 2002, \apj, 566, 123

\bibitem[{{Beers} {et~al.}(1990){Beers}, {Flynn}, \& {Gebhardt}}]{beers90}
{Beers}, T.~C., {Flynn}, K., \& {Gebhardt}, K. 1990, \aj, 100, 32

\bibitem[{{Bertin} \& {Arnouts}(1996)}]{bertin96}
{Bertin}, E. \& {Arnouts}, S. 1996, \aaps, 117, 393

\bibitem[{{Blakeslee} {et~al.}(2003){Blakeslee}, {Franx}, {Postman}, {Rosati},
  {Holden}, {Illingworth}, {Ford}, {Cross}, {Gronwall}, {Ben{\'{\i}}tez},
  {Bouwens}, {Broadhurst}, {Clampin}, {Demarco}, {Golimowski}, {Hartig},
  {Infante}, {Martel}, {Miley}, {Menanteau}, {Meurer}, {Sirianni}, \&
  {White}}]{blakeslee03}
{Blakeslee}, J.~P., {Franx}, M., {Postman}, M., {et~al.} 2003, \apjl, 596, L143

\bibitem[{{Bruzual} \& {Charlot}(2003)}]{bruzual03}
{Bruzual}, G. \& {Charlot}, S. 2003, \mnras, 344, 1000

\bibitem[{{Cardelli} {et~al.}(1989){Cardelli}, {Clayton}, \&
  {Mathis}}]{cardelli89}
{Cardelli}, J.~A., {Clayton}, G.~C., \& {Mathis}, J.~S. 1989, \apj, 345, 245

\bibitem[{{Carlberg} {et~al.}(1997){Carlberg}, {Yee}, \&
  {Ellingson}}]{carlberg97}
{Carlberg}, R.~G., {Yee}, H.~K.~C., \& {Ellingson}, E. 1997, \apj, 478, 462

\bibitem[{{Chabrier}(2003)}]{chabrier03}
{Chabrier}, G. 2003, \pasp, 115, 763

\bibitem[{{Charlot} \& {Fall}(2000)}]{charlot00}
{Charlot}, S. \& {Fall}, S.~M. 2000, \apj, 539, 718

\bibitem[{{Cooper} {et~al.}(2007){Cooper}, {Newman}, {Coil}, {Croton}, {Gerke},
  {Yan}, {Davis}, {Faber}, {Guhathakurta}, {Koo}, {Weiner}, \&
  {Willmer}}]{cooper07}
{Cooper}, M.~C., {Newman}, J.~A., {Coil}, A.~L., {et~al.} 2007, \mnras, 376,
  1445

\bibitem[{{Cowie} \& {Barger}(2008)}]{cowie08}
{Cowie}, L.~L. \& {Barger}, A.~J. 2008, ArXiv e-prints, 806

\bibitem[{{Cucciati} {et~al.}(2006){Cucciati}, {Iovino}, {Marinoni}, {Ilbert},
  {Bardelli}, {Franzetti}, {Le F{\`e}vre}, {Pollo}, {Zamorani}, {Cappi},
  {Guzzo}, {McCracken}, {Meneux}, {Scaramella}, {Scodeggio}, {Tresse}, {Zucca},
  {Bottini}, {Garilli}, {Le Brun}, {Maccagni}, {Picat}, {Vettolani},
  {Zanichelli}, {Adami}, {Arnaboldi}, {Arnouts}, {Bolzonella}, {Charlot},
  {Ciliegi}, {Contini}, {Foucaud}, {Gavignaud}, {Marano}, {Mazure}, {Merighi},
  {Paltani}, {Pell{\`o}}, {Pozzetti}, {Radovich}, {Bondi}, {Bongiorno},
  {Busarello}, {de La Torre}, {Gregorini}, {Lamareille}, {Mathez}, {Mellier},
  {Merluzzi}, {Ripepi}, {Rizzo}, {Temporin}, \& {Vergani}}]{cucciati06}
{Cucciati}, O., {Iovino}, A., {Marinoni}, C., {et~al.} 2006, \aap, 458, 39

\bibitem[{{De Lucia} {et~al.}(2007){De Lucia}, {Poggianti},
  {Arag{\'o}n-Salamanca}, {White}, {Zaritsky}, {Clowe}, {Halliday}, {Jablonka},
  {von der Linden}, {Milvang-Jensen}, {Pell{\'o}}, {Rudnick}, {Saglia}, \&
  {Simard}}]{delucia07}
{De Lucia}, G., {Poggianti}, B.~M., {Arag{\'o}n-Salamanca}, A., {et~al.} 2007,
  \mnras, 374, 809

\bibitem[{{Demarco} {et~al.}(2007){Demarco}, {Rosati}, {Lidman}, {Girardi},
  {Nonino}, {Rettura}, {Strazzullo}, {van der Wel}, {Ford}, {Mainieri},
  {Holden}, {Stanford}, {Blakeslee}, {Gobat}, {Postman}, {Tozzi}, {Overzier},
  {Zirm}, {Ben{\'{\i}}tez}, {Homeier}, {Illingworth}, {Infante}, {Jee}, {Mei},
  {Menanteau}, {Motta}, {Zheng}, {Clampin}, \& {Hartig}}]{demarco07}
{Demarco}, R., {Rosati}, P., {Lidman}, C., {et~al.} 2007, \apj, 663, 164

\bibitem[{{Diaferio} {et~al.}(1999){Diaferio}, {Kauffmann}, {Colberg}, \&
  {White}}]{diaferio99}
{Diaferio}, A., {Kauffmann}, G., {Colberg}, J.~M., \& {White}, S.~D.~M. 1999,
  \mnras, 307, 537

\bibitem[{{Elbaz} {et~al.}(2007){Elbaz}, {Daddi}, {Le Borgne}, {Dickinson},
  {Alexander}, {Chary}, {Starck}, {Brandt}, {Kitzbichler}, {MacDonald},
  {Nonino}, {Popesso}, {Stern}, \& {Vanzella}}]{elbaz07}
{Elbaz}, D., {Daddi}, E., {Le Borgne}, D., {et~al.} 2007, \aap, 468, 33

\bibitem[{{Finoguenov} {et~al.}(2007){Finoguenov}, {Guzzo}, {Hasinger},
  {Scoville}, {Aussel}, {B{\"o}hringer}, {Brusa}, {Capak}, {Cappelluti},
  {Comastri}, {Giodini}, {Griffiths}, {Impey}, {Koekemoer}, {Kneib},
  {Leauthaud}, {Le F{\`e}vre}, {Lilly}, {Mainieri}, {Massey}, {McCracken},
  {Mobasher}, {Murayama}, {Peacock}, {Sakelliou}, {Schinnerer}, {Silverman},
  {Smol{\v c}i{\'c}}, {Taniguchi}, {Tasca}, {Taylor}, {Trump}, \&
  {Zamorani}}]{finoguenov07}
{Finoguenov}, A., {Guzzo}, L., {Hasinger}, G., {et~al.} 2007, \apjs, 172, 182

\bibitem[{{Furusawa} {et~al.}(2000){Furusawa}, {Shimasaku}, {Doi}, \&
  {Okamura}}]{furusawa00}
{Furusawa}, H., {Shimasaku}, K., {Doi}, M., \& {Okamura}, S. 2000, \apj, 534,
  624

\bibitem[{{Gerke} {et~al.}(2007){Gerke}, {Newman}, {Faber}, {Cooper}, {Croton},
  {Davis}, {Willmer}, {Yan}, {Coil}, {Guhathakurta}, {Koo}, \&
  {Weiner}}]{gerke07}
{Gerke}, B.~F., {Newman}, J.~A., {Faber}, S.~M., {et~al.} 2007, \mnras, 376,
  1425

\bibitem[{{Gilbank} {et~al.}(2008){Gilbank}, {Yee}, {Ellingson}, {Gladders},
  {Loh}, {Barrientos}, \& {Barkhouse}}]{gilbank08}
{Gilbank}, D.~G., {Yee}, H.~K.~C., {Ellingson}, E., {et~al.} 2008, \apj, 673,
  742

\bibitem[{{Holden} {et~al.}(2004){Holden}, {Stanford}, {Eisenhardt}, \&
  {Dickinson}}]{holden04}
{Holden}, B.~P., {Stanford}, S.~A., {Eisenhardt}, P., \& {Dickinson}, M. 2004,
  \aj, 127, 2484

\bibitem[{{Jarrett} {et~al.}(2000){Jarrett}, {Chester}, {Cutri}, {Schneider},
  {Skrutskie}, \& {Huchra}}]{jarrett00}
{Jarrett}, T.~H., {Chester}, T., {Cutri}, R., {et~al.} 2000, \aj, 119, 2498

\bibitem[{{Kodama} {et~al.}(2005){Kodama}, {Tanaka}, {Tamura}, {Yahagi},
  {Nagashima}, {Tanaka}, {Arimoto}, {Futamase}, {Iye}, {Karasawa}, {Kashikawa},
  {Kawasaki}, {Kitayama}, {Matsuhara}, {Nakata}, {Ohashi}, {Ohta}, {Okamoto},
  {Okamura}, {Shimasaku}, {Suto}, {Tamura}, {Umetsu}, \& {Yamada}}]{kodama05}
{Kodama}, T., {Tanaka}, M., {Tamura}, T., {et~al.} 2005, \pasj, 57, 309

\bibitem[{{Koyama} {et~al.}(2007){Koyama}, {Kodama}, {Tanaka}, {Shimasaku}, \&
  {Okamura}}]{koyama07}
{Koyama}, Y., {Kodama}, T., {Tanaka}, M., {Shimasaku}, K., \& {Okamura}, S.
  2007, \mnras, 382, 1719

\bibitem[{{Lidman} {et~al.}(2004){Lidman}, {Rosati}, {Demarco}, {Nonino},
  {Mainieri}, {Stanford}, \& {Toft}}]{lidman04}
{Lidman}, C., {Rosati}, P., {Demarco}, R., {et~al.} 2004, \aap, 416, 829

\bibitem[{{Lidman} {et~al.}(2008){Lidman}, {Rosati}, {Tanaka}, {Strazzullo},
  {Demarco}, {Ageorges}, {Casali}, {Kissler-Patig}, {Petr-Gotzens}, \&
  {Selman}}]{lidman08}
{Lidman}, C., {Rosati}, P., {Tanaka}, M., {et~al.} 2008, \aap, accepted,

\bibitem[{{Madau}(1995)}]{madau95}
{Madau}, P. 1995, \apj, 441, 18

\bibitem[{{Mart{\'{\i}}nez} {et~al.}(2002){Mart{\'{\i}}nez}, {Zandivarez},
  {Dom{\'{\i}}nguez}, {Merch{\'a}n}, \& {Lambas}}]{martinez02}
{Mart{\'{\i}}nez}, H.~J., {Zandivarez}, A., {Dom{\'{\i}}nguez}, M.,
  {Merch{\'a}n}, M.~E., \& {Lambas}, D.~G. 2002, \mnras, 333, L31

\bibitem[{{Mei}(2008)}]{mei08}
{Mei}, S. 2008, \apj, submitted,

\bibitem[{{Mei} {et~al.}(2006{\natexlab{a}}){Mei}, {Blakeslee}, {Stanford},
  {Holden}, {Rosati}, {Strazzullo}, {Homeier}, {Postman}, {Franx}, {Rettura},
  {Ford}, {Illingworth}, {Ettori}, {Bouwens}, {Demarco}, {Martel}, {Clampin},
  {Hartig}, {Eisenhardt}, {Ardila}, {Bartko}, {Ben{\'{\i}}tez}, {Bradley},
  {Broadhurst}, {Brown}, {Burrows}, {Cheng}, {Cross}, {Feldman}, {Golimowski},
  {Goto}, {Gronwall}, {Infante}, {Kimble}, {Krist}, {Lesser}, {Menanteau},
  {Meurer}, {Miley}, {Motta}, {Sirianni}, {Sparks}, {Tran}, {Tsvetanov},
  {White}, \& {Zheng}}]{mei06a}
{Mei}, S., {Blakeslee}, J.~P., {Stanford}, S.~A., {et~al.} 2006{\natexlab{a}},
  \apj, 639, 81

\bibitem[{{Mei} {et~al.}(2006{\natexlab{b}}){Mei}, {Holden}, {Blakeslee},
  {Rosati}, {Postman}, {Jee}, {Rettura}, {Sirianni}, {Demarco}, {Ford},
  {Franx}, {Homeier}, \& {Illingworth}}]{mei06b}
{Mei}, S., {Holden}, B.~P., {Blakeslee}, J.~P., {et~al.} 2006{\natexlab{b}},
  \apj, 644, 759

\bibitem[{{Merch{\'a}n} \& {Zandivarez}(2002)}]{merchan02}
{Merch{\'a}n}, M. \& {Zandivarez}, A. 2002, \mnras, 335, 216

\bibitem[{{Mullis} {et~al.}(2005){Mullis}, {Rosati}, {Lamer}, {B{\"o}hringer},
  {Schwope}, {Schuecker}, \& {Fassbender}}]{mullis05}
{Mullis}, C.~R., {Rosati}, P., {Lamer}, G., {et~al.} 2005, \apjl, 623, L85

\bibitem[{{Nakata} {et~al.}(2005){Nakata}, {Kodama}, {Shimasaku}, {Doi},
  {Furusawa}, {Hamabe}, {Kimura}, {Komiyama}, {Miyazaki}, {Okamura}, {Ouchi},
  {Sekiguchi}, {Ueda}, {Yagi}, \& {Yasuda}}]{nakata05}
{Nakata}, F., {Kodama}, T., {Shimasaku}, K., {et~al.} 2005, \mnras, 357, 1357

\bibitem[{{Ramella} {et~al.}(1997){Ramella}, {Pisani}, \& {Geller}}]{ramella97}
{Ramella}, M., {Pisani}, A., \& {Geller}, M.~J. 1997, \aj, 113, 483

\bibitem[{{Rines} {et~al.}(2005){Rines}, {Geller}, {Kurtz}, \&
  {Diaferio}}]{rines05}
{Rines}, K., {Geller}, M.~J., {Kurtz}, M.~J., \& {Diaferio}, A. 2005, \aj, 130,
  1482

\bibitem[{{Rosati} {et~al.}(1998){Rosati}, {della Ceca}, {Norman}, \&
  {Giacconi}}]{rosati98}
{Rosati}, P., {della Ceca}, R., {Norman}, C., \& {Giacconi}, R. 1998, \apjl,
  492, L21+

\bibitem[{{Rosati} {et~al.}(2004){Rosati}, {Tozzi}, {Ettori}, {Mainieri},
  {Demarco}, {Stanford}, {Lidman}, {Nonino}, {Borgani}, {Della Ceca},
  {Eisenhardt}, {Holden}, \& {Norman}}]{rosati04}
{Rosati}, P., {Tozzi}, P., {Ettori}, S., {et~al.} 2004, \aj, 127, 230

\bibitem[{{Schlegel} {et~al.}(1998){Schlegel}, {Finkbeiner}, \&
  {Davis}}]{schlegel98}
{Schlegel}, D.~J., {Finkbeiner}, D.~P., \& {Davis}, M. 1998, \apj, 500, 525

\bibitem[{{Stanford} {et~al.}(2005){Stanford}, {Eisenhardt}, {Brodwin},
  {Gonzalez}, {Stern}, {Jannuzi}, {Dey}, {Brown}, {McKenzie}, \&
  {Elston}}]{stanford05}
{Stanford}, S.~A., {Eisenhardt}, P.~R., {Brodwin}, M., {et~al.} 2005, \apjl,
  634, L129

\bibitem[{{Stanford} {et~al.}(2002){Stanford}, {Holden}, {Rosati},
  {Eisenhardt}, {Stern}, {Squires}, \& {Spinrad}}]{stanford02}
{Stanford}, S.~A., {Holden}, B., {Rosati}, P., {et~al.} 2002, \aj, 123, 619

\bibitem[{{Stanford} {et~al.}(2006){Stanford}, {Romer}, {Sabirli}, {Davidson},
  {Hilton}, {Viana}, {Collins}, {Kay}, {Liddle}, {Mann}, {Miller}, {Nichol},
  {West}, {Conselice}, {Spinrad}, {Stern}, \& {Bundy}}]{stanford06}
{Stanford}, S.~A., {Romer}, A.~K., {Sabirli}, K., {et~al.} 2006, \apjl, 646,
  L13

\bibitem[{{Strauss} {et~al.}(2002){Strauss}, {Weinberg}, {Lupton}, {Narayanan},
  {Annis}, {Bernardi}, {Blanton}, {Burles}, {Connolly}, {Dalcanton}, {Doi},
  {Eisenstein}, {Frieman}, {Fukugita}, {Gunn}, {Ivezi{\'c}}, {Kent}, {Kim},
  {Knapp}, {Kron}, {Munn}, {Newberg}, {Nichol}, {Okamura}, {Quinn}, {Richmond},
  {Schlegel}, {Shimasaku}, {SubbaRao}, {Szalay}, {Vanden Berk}, {Vogeley},
  {Yanny}, {Yasuda}, {York}, \& {Zehavi}}]{strauss02}
{Strauss}, M.~A., {Weinberg}, D.~H., {Lupton}, R.~H., {et~al.} 2002, \aj, 124,
  1810

\bibitem[{{Strazzullo} {et~al.}(2006){Strazzullo}, {Rosati}, {Stanford},
  {Lidman}, {Nonino}, {Demarco}, {Eisenhardt}, {Ettori}, {Mainieri}, \&
  {Toft}}]{strazzullo06}
{Strazzullo}, V., {Rosati}, P., {Stanford}, S.~A., {et~al.} 2006, \aap, 450,
  909

\bibitem[{{Tanaka} {et~al.}(2004){Tanaka}, {Goto}, {Okamura}, {Shimasaku}, \&
  {Brinkmann}}]{tanaka04}
{Tanaka}, M., {Goto}, T., {Okamura}, S., {Shimasaku}, K., \& {Brinkmann}, J.
  2004, \aj, 128, 2677

\bibitem[{{Tanaka} {et~al.}(2007{\natexlab{a}}){Tanaka}, {Hoshi}, {Kodama}, \&
  {Kashikawa}}]{tanaka07b}
{Tanaka}, M., {Hoshi}, T., {Kodama}, T., \& {Kashikawa}, N. 2007{\natexlab{a}},
  \mnras, 379, 1546

\bibitem[{{Tanaka} {et~al.}(2005){Tanaka}, {Kodama}, {Arimoto}, {Okamura},
  {Umetsu}, {Shimasaku}, {Tanaka}, \& {Yamada}}]{tanaka05}
{Tanaka}, M., {Kodama}, T., {Arimoto}, N., {et~al.} 2005, \mnras, 362, 268

\bibitem[{{Tanaka} {et~al.}(2006){Tanaka}, {Kodama}, {Arimoto}, \&
  {Tanaka}}]{tanaka06}
{Tanaka}, M., {Kodama}, T., {Arimoto}, N., \& {Tanaka}, I. 2006, \mnras, 365,
  1392

\bibitem[{{Tanaka} {et~al.}(2007{\natexlab{b}}){Tanaka}, {Kodama}, {Kajisawa},
  {Bower}, {Demarco}, {Finoguenov}, {Lidman}, \& {Rosati}}]{tanaka07a}
{Tanaka}, M., {Kodama}, T., {Kajisawa}, M., {et~al.} 2007{\natexlab{b}},
  \mnras, 377, 1206

\bibitem[{{Thomas} {et~al.}(2005){Thomas}, {Maraston}, {Bender}, \& {Mendes de
  Oliveira}}]{thomas05}
{Thomas}, D., {Maraston}, C., {Bender}, R., \& {Mendes de Oliveira}, C. 2005,
  \apj, 621, 673

\bibitem[{{Tran} {et~al.}(2001){Tran}, {Simard}, {Zabludoff}, \&
  {Mulchaey}}]{tran01}
{Tran}, K.-V.~H., {Simard}, L., {Zabludoff}, A.~I., \& {Mulchaey}, J.~S. 2001,
  \apj, 549, 172

\bibitem[{{Umetsu} {et~al.}(2005){Umetsu}, {Tanaka}, {Kodama}, {Tanaka},
  {Futamase}, {Kashikawa}, \& {Hoshi}}]{umetsu05}
{Umetsu}, K., {Tanaka}, M., {Kodama}, T., {et~al.} 2005, \pasj, 57, 877

\bibitem[{{Yagi} {et~al.}(2002){Yagi}, {Kashikawa}, {Sekiguchi}, {Doi},
  {Yasuda}, {Shimasaku}, \& {Okamura}}]{yagi02}
{Yagi}, M., {Kashikawa}, N., {Sekiguchi}, M., {et~al.} 2002, \aj, 123, 66

\bibitem[{{York} {et~al.}(2000){York}, {Adelman}, {Anderson}, {Anderson},
  {Annis}, {Bahcall}, {Bakken}, {Barkhouser}, {Bastian}, {Berman}, {Boroski},
  {Bracker}, {Briegel}, {Briggs}, {Brinkmann}, {Brunner}, {Burles}, {Carey},
  {Carr}, {Castander}, {Chen}, {Colestock}, {Connolly}, {Crocker}, {Csabai},
  {Czarapata}, {Davis}, {Doi}, {Dombeck}, {Eisenstein}, {Ellman}, {Elms},
  {Evans}, {Fan}, {Federwitz}, {Fiscelli}, {Friedman}, {Frieman}, {Fukugita},
  {Gillespie}, {Gunn}, {Gurbani}, {de Haas}, {Haldeman}, {Harris}, {Hayes},
  {Heckman}, {Hennessy}, {Hindsley}, {Holm}, {Holmgren}, {Huang}, {Hull},
  {Husby}, {Ichikawa}, {Ichikawa}, {Ivezi{\'c}}, {Kent}, {Kim}, {Kinney},
  {Klaene}, {Kleinman}, {Kleinman}, {Knapp}, {Korienek}, {Kron}, {Kunszt},
  {Lamb}, {Lee}, {Leger}, {Limmongkol}, {Lindenmeyer}, {Long}, {Loomis},
  {Loveday}, {Lucinio}, {Lupton}, {MacKinnon}, {Mannery}, {Mantsch}, {Margon},
  {McGehee}, {McKay}, {Meiksin}, {Merelli}, {Monet}, {Munn}, {Narayanan},
  {Nash}, {Neilsen}, {Neswold}, {Newberg}, {Nichol}, {Nicinski}, {Nonino},
  {Okada}, {Okamura}, {Ostriker}, {Owen}, {Pauls}, {Peoples}, {Peterson},
  {Petravick}, {Pier}, {Pope}, {Pordes}, {Prosapio}, {Rechenmacher}, {Quinn},
  {Richards}, {Richmond}, {Rivetta}, {Rockosi}, {Ruthmansdorfer}, {Sandford},
  {Schlegel}, {Schneider}, {Sekiguchi}, {Sergey}, {Shimasaku}, {Siegmund},
  {Smee}, {Smith}, {Snedden}, {Stone}, {Stoughton}, {Strauss}, {Stubbs},
  {SubbaRao}, {Szalay}, {Szapudi}, {Szokoly}, {Thakar}, {Tremonti}, {Tucker},
  {Uomoto}, {Vanden Berk}, {Vogeley}, {Waddell}, {Wang}, {Watanabe},
  {Weinberg}, {Yanny}, \& {Yasuda}}]{york00}
{York}, D.~G., {Adelman}, J., {Anderson}, Jr., J.~E., {et~al.} 2000, \aj, 120,
  1579

\bibitem[{{Zabludoff} \& {Mulchaey}(1998)}]{zabludoff98}
{Zabludoff}, A.~I. \& {Mulchaey}, J.~S. 1998, \apj, 496, 39

\end{thebibliography}

\end{document}